\documentclass[12pt]{article}

\usepackage{graphicx}
\usepackage{amssymb}
\usepackage{amsfonts}
\usepackage{comment}

\usepackage{amsmath,amsthm,amsfonts,amscd} 
\usepackage{verbatim}           
\usepackage{psfig}              
\usepackage{epsfig}             
\usepackage{fullpage}
\usepackage{comment}
\usepackage{mathrsfs}

\newtheorem{preremark}{Remark}[section]
\newenvironment{rem}{\begin{preremark}\rm}{\end{preremark}}
\newtheorem{defi}{Definition}[section]
\newtheorem{theo}{Theorem}[section]

\newtheorem{cor}{Corollary}[section]
\newtheorem{conj}{Conjecture}[section]

\newcommand{\Author}[1]{{#1},}
\newcommand{\ArticleTitle}[1]{``{#1}'',}
\newcommand{\BookTitle}[1]{{\em #1},}
\newcommand{\JournalTitle}[1]{{\em #1}}

\newcommand{\regul}{{\kappa}}

\newcommand{\rarefied}{{\theta_{\mathrm{rar}}}}
\newcommand{\dense}{{\theta_{\mathrm{den}}}}

\newcommand{\golden}{{\sigma_{\mathrm{G}}}}

\newcommand{\calV}{{\mathcal{V}}}

\newcommand{\calX}{{\mathcal{X}}}

\newcommand{\bbN}{{\mathbb{N}}}

\newcommand{\bbR}{{\mathbb{R}}}

\newcommand{\bbT}{{\mathbb{T}}}

\newcommand{\bbZ}{{\mathbb{Z}}}

\begin{document}

\title{Regularity Properties 
	of Critical Invariant Circles \\
	of Twist Maps, and Their Universality}
\author{
Arturo Olvera$^1$\thanks{
E-mail:  \tt aoc@uxmym1.iimas.unam.mx
}
\and
Nikola P. Petrov$^{2}$\thanks{
E-mail:  \tt npetrov@ou.edu
}
}
\date{\today}

\maketitle

\begin{center}$^1$
IIMAS-UNAM, FENOMEC, Apdo.\ Postal 20--726, 
M\'exico D.\ F.\ 01000, Mexico\\[6mm]
\end{center}

\begin{center}$^2$
Department of Mathematics, University of Oklahoma, 
Norman, OK 73019, USA
\end{center}

\begin{abstract}
We compute accurately the golden critical invariant circles 
of several area-preserving twist maps of the cylinder.  
We define some functions related to the invariant 
circle and to the dynamics of the map restricted to the circle 
(for example, the conjugacy between the circle map 
giving the dynamics on the invariant circle 
and a rigid rotation on the circle).  
The global H\"older regularities of these functions are low 
(some of them are not even once differentiable).  
We present several conjectures about the universality 
of the regularity properties of the critical circles 
and the related functions.  
Using a Fourier analysis method 
developed by R.\ de la Llave and one of the authors, 
we compute numerically the H\"older regularities of these functions.  
Our computations show that -- withing their numerical accuracy -- 
these regularities are the same for the different maps studied.  
We discuss how our findings are related to some previous results:  
(a) to the constants giving the scaling behavior of the iterates 
on the critical invariant circle (discovered by Kadanoff and Shenker); 
(b) to some characteristics of the singular invariant measures 
connected with the distribution of iterates.  
Some of the functions studied have pointwise H\"older 
regularity that is different at different points.  
Our results give a convincing numerical support 
to the fact that the points with different H\"older exponents 
of these functions are interspersed in the same way for 
different maps, which is a strong indication 
that the underlying twist maps belong to the same universality class.  
\end{abstract}


\pagebreak




{\bf 
Area-preserving twist maps of the cylinder 
are popular models of the dynamics 
of many physical systems, 
i.e., they occur as Poincar\'e maps 
of 2-degree-of-freedom Hamiltonian systems.  
Homotopically non-trivial invariant circles 
of such a map play an important role 
in organizing the global dynamics of the map.  
Generally, as the perturbation grows, 
more and more of these circles are destroyed, 
until there remains only one such circle, 
called the ``critical'' circle.  
This circle is the last obstacle to the unbounded growth 
of the ``action variable''.  
In this critical situation many characteristics 
of the system become drastically different 
from the ``under-critical'' case.  
For example, consider the dynamics of the iterates 
of the twist map restricted to the critical circle 
-- it is given by a map of the circle.  
This map can be conjugated to a rigid rotation 
on the circle, but the conjugating function 
has very low regularity -- 
its H\"older exponent is lower than~1.  
The H\"older regularity of this conjugacy 
is related to some universal properties of the map, 
i.e., to the universal rescaling factors 
\cite{Kadanoff81,ShenkerK82} 
and to the scaling properties \cite{HalseyJKPS86} 
of the distribution 
of the iterates on the critical circle 
(which is governed by a singular invariant measure).  
We compute several functions related 
to the critical circle and to the dynamics on it 
and us a method developed in \cite{LlaveP02} 
to assess numerically their global H\"older regularity.  
Our findings lend support to several conjectures 
concerned with the universality and the renormalization 
group description of these critical objects.

}

\bigskip


\section{Introduction}\label{sec:intro}

It has been known since the late 1970's and early 1980's 
that many objects at the boundary of chaotic behavior 
exhibit remarkable scaling properties 
and that, furthermore, these properties are ``universal''.  
Such properties are exhibited by unimodal maps of the interval 
\cite{TresserC78,Feigenbaum78,Feigenbaum79}, 
critical maps of the circle \cite{Shenker82}, 
critical KAM tori \cite{Kadanoff81,ShenkerK82}, and other systems.  
These observations were explained in terms of a renormalization 
group analysis, following a methodology that had been 
developed earlier in the study of critical phenomena 
in statistical mechanics and field theory 
\cite{ColletEL80,FKS82,ORSS83,MacKay83}.  

The scale invariance of the critical objects 
affects many of their properties.  
Notably, the H\"older regularity 
of the critical objects (or some functions related to them) 
tends to have a low and fractional value.  
Presumably the values of the regularities 
are related to exponents and geometric properties 
of the renormalization group fixed points 
which describe the critical objects.  

Furthermore, the observation 
that critical objects can be divided 
in ``universality classes" such that all objects in a given class 
``look the same'' can be tested numerically.  
One way to do this is to define 
certain functions related to the critical objects 
-- typically these functions are not very regular 
(in some cases not even once differentiable), 
and to test numerically whether the regularities 
of these functions are the same for different objects.  
Another -- even more sensitive -- test for universality 
is to take two functions, say $h_1$ and $h_2$, 
from the same class, and to study the regularity 
of functions like $h_1\circ h_2^{-1}$ 
-- for $h_1$ and $h_2$ belonging to the same universality class, 
one can expect that $h_1\circ h_2^{-1}$ 
be more regular than $h_1$ and~$h_2^{-1}$.  
These ideas were tested in \cite{LlaveP02} 
in the case of non-critical and critical 
(with different degree of criticality) circle maps, 
in which the empirical results are accompanied 
by an extensive mathematical theory.  
A substantial part of the effort in \cite{LlaveP02} 
was to develop implementations of methods 
known in harmonic analysis 
(finite differences, Littlewood-Paley theory, wavelet analysis) 
to assess the regularity of the objects numerically.

In the present paper, we extend the methodology 
of \cite{LlaveP02} to the study of critical invariant 
circles of area-preserving twist maps.  
Invariant circles in dynamical systems 
are among the most important objects 
that organize the long-term behavior of the system, 
and the critical ones are especially important 
because of their role as ``last barriers to chaos'' 
(for readable reviews see, e.g.,~\cite{Meiss92}, 
or, with more emphasis on the mathematical aspects, 
the recent book~\cite{Gole01}).  
Critical invariant circles have been extensively studied since 
early 1980's \cite{Kadanoff81,ShenkerK82,MacKay82,MacKay83}.  

We compute accurately the golden critical invariant 
circles of several standard-like area-preserving twist maps 
and some functions related to the dynamics 
of the iterates of the maps on these circles.  
Then we apply methods developed in \cite{LlaveP02} 
to study the H\"older regularity of these functions 
and some universality aspects.  

In Sec.~\ref{sec:background} 
we give some background on twist maps and their 
critical invariant circles, 
define the functions that are the objects 
of our numerical study, 
and present several precise conjectures 
concerning the properties of the critical invariant circles  
and the functions introduced.  
Sec.~\ref{sec:numerical-methods} 
is devoted to a discussion of the numerical methods used 
to compute critical invariant circles 
and to assess H\"older regularity of functions.  
We collect our results in Sec.~\ref{sec:numerical-results}, 
and in Sec.~\ref{sec:discussion} 
we discuss their significance and relationship 
with previous studies.


\section{\label{sec:background}Critical invariant circles of twist maps}   



\subsection{\label{sec:twist-maps}Twist maps}

Let $\bbT:=\bbR/\bbZ$ stand for the circle.  
We will be concerned with maps 
$F$ of the (infinite) cylinder $\bbT \times \bbR$, 
\[
F : \bbT \times \bbR \to \bbT \times \bbR 
  : (\theta,r) \mapsto F(\theta,r) 
        =: (\theta',r') \ ,
\]
which satisfy the following properties:  
\begin{itemize}
\item
{\em Area preservation:} 
The map $F$ preserves the oriented area:  
$\det DF = 1$.  

\item 
{\em Zero-flux:}  
The oriented area between a homotopically non-trivial circle 
and its image under $F$ is~$0$.  
(In our situation, this is equivalent to saying that 
every non-trivial circle intersects its image.)  

\item
{\em Twist condition:}  
For any fixed value of~$\theta$, 
$\frac{\partial \theta'}{\partial r}  > 0$.  

\end{itemize}

A map of the cylinder can be identified 
with a map $\tilde F:\bbR^2\to\bbR^2$ 
(called a {\em lift} of~$F$) which satisfies 
\[
\tilde F(\theta+1,r) = \tilde F(\theta,r) + (1,0)  \ .
\]
Often one does not need to keep the distinction.  


The maps which we will use in our numerical studies 
are of the form $(\theta',r') := F(\theta,r)$ with 
\begin{eqnarray}  \label{eq:standard}
\theta' &=& (\theta + r') \ \mathrm{mod}\,1  \ , \nonumber \\[-2mm]
\\[-2mm]
r' &=& r + \lambda \, V(\theta) \nonumber \ , \nonumber
\end{eqnarray}
where $\lambda$ is a parameter, 
and $V:\bbT \to \bbR$ is a function 
satisfying $\int_0^1 V(\theta) \, d\theta = 0$.  
In particular, many numerical studies 
have been devoted to studying \eqref{eq:standard} with 
\begin{equation}  \label{eq:Vsin}
V(\theta) = - \frac1{2\pi} \, \sin 2 \pi \theta \ ,
\end{equation}  
in which case we will call the map $F$ 
the {\em Taylor-Chirikov} map.

Given an orbit $\calX=
\left\{(\theta_n,r_n) = F^n(\theta_0,r_0) 
\, | \, n=0,1,2,\ldots \right\}$, 
we define its {\em rotation number}, $\rho(\calX)$, as the limit 
\[
\rho(\calX) := \lim_{n\to\pm\infty} \frac{\theta_n-\theta_0}{n} \ ,
\]
whenever this limit exists.  
In contrast with the situation for circle maps, 
the rotation number depends on the orbit 
(and it may happen that some orbits 
do not have rotation number).  

We say that an orbit is {\em well-ordered} 
when for every $k$ and $l$, the function of $n$ defined as 
$e(n) = \theta_{n+k} - l - \theta_n$ 
has the same sign.  
Every well-ordered orbit has a rotation number 
(the converse, however, is not true).  

It is also easy to see that if a bounded orbit 
is well-ordered and $\rho(\theta_0,r_0)$ is irrational, 
the closure of the orbit, 
$\overline{\{(\theta_n,r_n)\}_{n=0}^\infty}$, 
is a perfect set 
(i.e., every point is an accumulation point 
of points in the set); 
in other words, in this case 
the orbit is either a homotopically non-trivial 
circle or a Cantor set.  

A set $U\subseteq \bbT \times \bbR$ is {\em invariant} 
if $U=F(U)$.  

The following result plays an important role.

\begin{theo}  \label{th:exist-orbit}
If $F$ is as above, 
for every $\rho \in \bbR$ there exists a well-ordered orbit 
with rotation number~$\rho$.  
\end{theo}


\subsection{\label{sec:rigorous}Invariant circles of twist maps 
-- rigorous results}

The proof of the following theorem can be found in 
\cite{Birkhoff20}, \cite{Mather84}.    
We refer to \cite{Gole01} and \cite{MatherF94} 
for a detailed exposition.  

\begin{theo}  \label{th:mather}
Let $U$ be an open simply connected 
invariant set, containing one of the ends of the cylinder.  
Then the boundary, $\partial U$, 
of the set $U$ is an invariant circle 
which is the graph of a Lipschitz function.  
In other words, $\partial U$ can be written as $r=R(\theta)$, 
where $R:\bbT\to\bbR$ is a Lipschitz function.  

For the map \eqref{eq:standard}, 
the Lipschitz constant of the function $R$ 
can be bounded by an expression which involves 
only the Lipschitz constant of the function~$F$ 
in a neighborhood of the circle~$\partial U$.  
\end{theo}

In particular,  

\begin{cor}
Any homotopically non-trivial invariant circle is the graph 
of a Lipschitz function~$R$.  
In the particular case of the map \eqref{eq:standard}, 
the Lipschitz constant of $R$ 
can be bounded by a constant which is independent 
of~$\lambda\,\mathrm{Lip}\,V$.  
\end{cor}


A number $\rho$ is said to be {\em Diophantine} 
if, for each $m,n\in\bbN\setminus\{0\}$, for some $C>0$, 
and for some~$d>2$, it satisfies 
\[
\left|\rho - \frac mn \right| > \frac C{n^d} \ .
\]

In the case when the map $F$ is close to integrable 
and its rotation number $\rho$ is Diophantine, 
one can apply Kolmogorov-Arnold-Moser theory to obtain 
that there exists an analytic invariant circle 
such that the orbits on it have 
rotation number~$\rho$.  

%
%

{\em Golden} invariant circles are those with rotation number 
equal to the {\em golden mean}, 
\begin{equation}  \label{eq:sigmaG}
\sigma_\mathrm{G} := [1,1,1,\ldots] = \frac{\sqrt{5}-1}2 \ .
\end{equation}
Here we have used the notation 
$\rho
= [a_1,a_2,a_3,\ldots] 
= 1/(a_1 + 1/(a_2 + 1/(a_3 + \cdots)))$
for the continued fraction expansion of $\rho\in(0,1)$ 
\cite{Khin97}.

There are also rigorous results that guarantee 
the non-existence of invariant circles 
of $F$ of the form~\eqref{eq:standard}.  

\begin{theo}   \label{th:no-circles}
\begin{itemize}
\item[(i)] 
If $\sup_\theta|\lambda\,V(\theta)| > 1$, 
then \eqref{eq:standard} has no invariant circles.  

\item[(ii)]
If $\sup_\theta|V'(\theta)| = 1$ 
(which holds for the function \eqref{eq:Vsin}), 
then for $|\lambda| > \frac43$ 
the map \eqref{eq:standard} has no invariant circles.  

\item[(iii)] 
For $V$ given by \eqref{eq:Vsin}, 
the map \eqref{eq:standard} has no golden invariant circles 
for $|\lambda| > \frac{63}{64}=0.984375\ldots$.  

\item[(iv)] 
For $V$ given by \eqref{eq:Vsin}, the map \eqref{eq:standard} 
has no golden invariant circles for $|\lambda| > 0.9718$.  
\end{itemize}
\end{theo}

Part (i) of Theorem \ref{th:no-circles} is elementary: 
if $\lambda\,\sup_\theta|V(\theta)| > 1$, 
then there will exist 
points $(\theta^*, r^*)\in\bbT \times \bbR$ 
such that 
$F(\theta^*, r^*) = (\theta^*, r^* + 1)$, 
which, iterated, gives that 
$F^n(\theta^*, r^*) = (\theta^*, r^* + n)$ 
-- the unbounded growth of the second coordinate 
of $F^n(\theta^*, r^*)$ with $n$ 
implies that a topologically non-trivial 
invariant circle cannot exist.  

Part (ii) can be found 
in \cite{Mather84}, 
parts (iii) and (iv) are proved 
by computer-assisted methods 
in \cite{MacKayP87} and \cite{Jungreis91}, 
resp.  

It is widely believed that 

\begin{conj} \label{conj:cotas}
For a Diophantine number $\rho$ 
and for a map $F$ of the form \eqref{eq:standard}, 
there is a number $\Lambda(\rho)$ such that 
when $|\lambda| > \Lambda(\rho)$, 
there is no invariant circle with rotation number $\rho$, 
and 
when $|\lambda| < \Lambda(\rho)$, 
there exists an analytic invariant circle 
with rotation number $\rho$. 
The invariant circle becomes critical when 
$|\lambda| = \Lambda(\rho)$.  
\end{conj}


Since our paper is devoted to homotopically non-trivial 
invariant circles, we will usually 
omit the words ``homotopically non-trivial''.


\subsection{\label{sec:functions}Functions related to the critical invariant circles}

We are interested in describing the critical invariant circles 
with rotation number $\rho$ which are in the boundary 
of existence.  
Postponing for the moment issues on how these objects 
can be actually computed, we point out that 
to a given critical invariant circle $\gamma$ 
of rotation number $\rho$, we can associate: 
\begin{itemize}
\item
the function $R:\bbT\to\bbR$ such that 
the critical invariant circle 
$\gamma$ is the graph of~$R$: 
\begin{equation}  \label{eq:R-def}
\gamma = \{(\theta,r)\in\bbT \times \bbR\,:\,r=R(\theta)\} \ ;
\end{equation}

\item
the {\em advance map} $g:\bbT\to\bbT$ defined by 
\begin{equation}  \label{eq:g-def}
F(\theta,R(\theta)) = (g(\theta), R\circ g(\theta)) \ ;
\end{equation}

\item
the {\em hull map} $\Psi:\bbT\to\bbT\times\bbR$, 
which gives a representation of the invariant circle $\gamma$ 
in such a way that 
the dynamics on $\gamma$ becomes a rotation by $\rho$, i.e., 
\begin{equation}  \label{eq:Psi-def}
F \circ \Psi(\theta) = \Psi (\theta + \rho)  \ ;
\end{equation}

\item 
the map $h = \pi_1 \circ \Psi: \bbT\to\bbT$ 
(where $\pi_1:\bbT\times\bbR\to\bbT$ is the projection onto $\bbT$), 
which conjugates the advance map to a rotation by $\rho$:  
\begin{equation}  \label{eq:h-def}
g \circ h(\theta) = h(\theta + \rho) \ ;
\end{equation}

\item
the map $h^{-1} : \bbT \to \bbT$, 
which is the inverse of the map~$h$ 
defined in~\eqref{eq:h-def}.  
\end{itemize}

We note the following rigorous results.  

Theorem \ref{th:mather} guarantees that 
the invariant circle the function $R$ is Lipschitz.  
It is an easy consequence of the implicit function theorem 
that $g$ should be as regular as $R$.  
Nevertheless, it is useful to compute the regularities 
of both $g$ and $R$ independently 
to asses the reliability of the numerical methods used.  

Because of \eqref{eq:h-def}, it is clear 
that the regularity of $g$ is not smaller than the minimum 
of the regularities of $h$ and~$h^{-1}$.  





\subsection{\label{sec:big-conj}The ``big'' conjugacies}


Let $\rho$ be a Diophantine number, 
$F_i$ ($i=1,2$) be area-preserving twist maps, 
and $\gamma_i$ be the critical invariant circle 
of $F_i$ with rotation number $\rho$.  
Let $g_{\gamma_i}$ and $h_{\gamma_i}$ 
be the associated advance map \eqref{eq:g-def} 
and conjugacy \eqref{eq:h-def}, resp.  
We introduce the conjugating functions 
\begin{eqnarray*}
G_{\gamma_1,\gamma_2} 
	\!&:=&\! g_{\gamma_1} \circ g_{\gamma_2}^{-1} 
	: \bbT \to \bbT \ , 
	\label{eq:G-def}   \\[1mm]
H_{\gamma_1,\gamma_2} 
	\!&:=&\! h_{\gamma_1} \circ h_{\gamma_2}^{-1} 
	: \bbT \to \bbT \ .
	\label{eq:H-def}   
\end{eqnarray*}
We will call these functions ``big'' conjugacies 
to distinguish them from the ``small'' conjugacies $h$ 
that conjugate the projected dynamics 
on the critical circles to a rigid rotation~\eqref{eq:h-def}.  
Note that the ``big'' conjugacies satisfy 
\[
G_{\gamma_1,\gamma_2} 
	\circ G_{\gamma_2,\gamma_3} 
	= G_{\gamma_1,\gamma_3} 
\ , \quad 
H_{\gamma_1,\gamma_2} 
	\circ H_{\gamma_2,\gamma_3} 
	= H_{\gamma_1,\gamma_3} \ .
\]
Below we discuss one aspect 
of the definition of the big conjugacies 
that will be important in our computations.  

Since there is no ``origin'' on the circle $\bbT$, 
one has certain amount of freedom 
in the definition of some maps.  
For example, if the function $\Psi$ is a hull map 
(i.e., satisfies \eqref{eq:Psi-def}), 
then the function $\tilde \Psi$ defined as 
$\tilde \Psi(\theta) = \Psi(\theta+\zeta)$ 
will also satisfy \eqref{eq:Psi-def} 
for any choice of the constant~$\zeta$.  
Similarly, the map $h$ \eqref{eq:h-def} 
that conjugates the advance map $g$ 
to a rigid rotation 
can be redefined by composing it on the right 
with a rotation, and the resulting map, 
$\tilde h(\theta) = h(\theta + \zeta)$, 
will also conjugate $g$ to a rigid rotation.  
Naturally, all important properties of 
the maps $h$ and $\tilde h$ 
-- in particular, their H\"older regularity
-- will be the same.  
However, one cannot use this freedom liberally 
when studying the big conjugacies.  
To understand the reason for this, 
consider the map $h$ defined by \eqref{eq:h-def}  
for some twist map~$F$.  
Naturally, the map $h\circ h^{-1}$ 
is the identity map, so it is~$C^\infty$.  
However, for any nonzero $\zeta$ 
in the definition of $\tilde h$, 
there is no guarantee 
that the map $h\circ \tilde h^{-1}$ 
will be~$C^\infty$.  
This is due to the fact that the regularity 
of $h$ may be different at different points, 
and while in $h\circ h^{-1}$ these ``irregularities'' 
cancel out, in $h\circ \tilde h^{-1}$ 
the action of $h$ does not necessarily ``undo'' 
the irregularities caused by $\tilde h^{-1}$.  
In Sec.~\ref{sec:H-sym} we explain 
in detail how we choose $\zeta$ 
in order to avoid the ``spurious'' irregularities 
of the big conjugacy.


\subsection{\label{sec:H-sym}Big conjugacies and symmetries}

Consider two functions 
$h_{\gamma_1}$ and $h_{\gamma_2}$ 
corresponding to the critical circles 
$\gamma_1$ and $\gamma_2$ 
of the twist maps $F_1$ and~$F_2$.  
If $F_1$ and $F_2$ 
happen to belong to the same 
``universality class'' 
(see Sec.~\ref{sec:universality}), 
then one would expect that the big conjugacy 
$H_{\gamma_1,\gamma_2}$ will be more regular than 
the functions $h_{\gamma_1}$ and~$h_{\gamma_2}^{-1}$.  
To avoid introducing spurious irregularities 
in $H_{\gamma_1,\gamma_2}$, 
we use the symmetries of the map $h$ 
that come from the symmetries 
of the~$F$~\cite{DeV58,Greene79,ApteLP05}.  

It is well known that if the function $V$ is odd, 
then the map $F$ given by \eqref{eq:standard} 
can be written as a composition of two involutions: 
\begin{equation}  \label{eq:Fdecomposed}
F = I_1 \circ I_0 \ , \qquad I_0^2 = I_1^2 = \mathrm{Id} \ ,
\end{equation}
where 
\begin{equation}  \label{eq:involutions}
I_0(\theta,r) = (-\theta, r+\lambda V(\theta)) \ , 
\quad 
I_1(\theta,r) = (-\theta + r, r) \ . 
\end{equation}
From \eqref{eq:Fdecomposed} 
we have 
$I_0 \circ F = F^{-1} \circ I_0$ 
and 
$I_1 \circ F = F^{-1} \circ I_1$.  
Acting on \eqref{eq:Psi-def} with $I_0$ from the left, 
we obtain 
\begin{equation}  \label{eq:compare1}
F^{-1}\circ (I_0 \circ \Psi)(\theta) 
= 
(I_0 \circ \Psi)(\theta+\rho) \ .
\end{equation}
On the other hand, if we define 
the function $L:\bbT\to\bbT\times\bbR$ by 
$L(\theta) := \Psi(-\theta)$, 
then \eqref{eq:Psi-def} can be written as 
\begin{equation}  \label{eq:compare2}
F^{-1}\circ L(\theta) = L(\theta+\rho) \ .
\end{equation}
Comparing \eqref{eq:compare1} and \eqref{eq:compare2}, 
we see that $L$ and $I_0\circ\Psi$ 
can differ only by a shift in the argument, 
i.e., there has to exist a constant $\zeta$ 
such that 
$I_0\circ\Psi(\theta) = L(\theta + \zeta) = \Psi(-\theta - \zeta)$.  
This, together with \eqref{eq:involutions} 
and $h=\pi_1\circ\Psi$, imply 
\[
h(\theta) = - h(-\theta - \zeta) \ .
\]
This implies that $h(-\frac{\zeta}{2})=0$, 
and the numerical value of $\zeta$ 
can be found from the computed values of~$h$.  
Setting 
$\tilde h (\theta) := h(\theta - \frac{\zeta}{2})$, 
we obtain that $\tilde h$ is an odd function.  
In what follows, we will assume that 
the appropriate value of $\zeta$ 
has been subtracted, and will omit 
the tilde over~$h$.


\subsection{\label{sec:universality}Universality}

In this section, we formulate precisely some conjectures 
on the behavior of critical invariant circles 
described by a non-trivial fixed point 
of the renormalization group.  
It seems quite possible that these conjectures 
can be proved as conditional theorems 
assuming existence and certain properties of this fixed point.  

One of the most striking predictions 
of the renormalization group theory is that many characteristics 
of the critical invariant circles are largely independent 
of the details of the map.  
This is captured by the notion of universality.  

\begin{defi}   \label{def:universality}
We say that a numerical characteristic is {\em universal} 
when it takes the same value 
in an open set of functions.  
We say that a property is {\em universal} 
when it holds for an open set of functions.  
\end{defi}

The open sets alluded to in Definition~\ref{def:universality} 
are called {\em domains of universality}.  


For the case that we will be concerned with, 
the description of the domains 
of universality in terms of properties 
of the non-trivial fixed points of the renormalization operator 
is still debated, but there are indications 
that the domain of universality is not the whole space 
\cite{Ketoja92,KetojaM94,FalcoliniL92-domains}.  

\begin{conj}   \label{conj:critical}
The existence of one and only one 
non-trivial fixed point of the renormalization operator 
is a universal property.  
\end{conj}

This conjecture has been known for a long time 
\cite{MacKay83}.  
Recently in \cite{LlaveO06} it has been shown 
how this conjecture follows rigorously 
from an extension of the standard renormalization 
group picture.  
Even the formulation of the subsequent conjectures 
depends of Conjecture~\ref{conj:critical}.  

%
%
%
%

The concept of universality is rather natural 
when one wants to study properties that depend on the speed 
at which the set of maps converges to the fixed point 
under the renormalization operator.  
In particular, 
regularity of conjugacies depends 
on the this speed of convergence 
and, hence, should be a universal quantity 
(more precise formulations are given in \cite{LlaveS96}).  
Hence, we can also conjecture that:  

\begin{conj}   \label{conj:regularity}
The regularity, $\regul(R)$, 
of the critical invariant circle is a universal number.  
\end{conj}

\begin{conj} \label{conj:stratified}
The regularities 
$\regul(g)$, $\regul(h)$ and $\regul(h^{-1})$ 
are universal numbers.  
\end{conj}

\begin{conj} \label{conj:big-conj}
For pairs of critical circles $\gamma_1$ and $\gamma_2$, 
the regularities 
$\regul(G_{\gamma_1,\gamma_2})$ 
and 
$\regul(H_{\gamma_1,\gamma_2})$ 
are universal numbers.  
\end{conj}




Directly from the definition of H\"older regularity, 
one can see that if 
$\regul(\phi)$ and $\regul(\psi)$ 
are between 0 and~1, 
then 
$\regul(\phi\circ\psi) \geq \regul(\phi)\,\regul(\psi)$.  
This implies that 
\begin{equation}  \label{eq:holder-comp}
\regul(H_{\gamma_1,\gamma_2}) 
= \regul(h_{\gamma_1}\circ h_{\gamma_2}^{-1}) 
\geq 
\regul(h_{\gamma_1})\, \regul(h_{\gamma_2}^{-1}) \ .
\end{equation}

For all critical invariant circles $\gamma_i$ 
that we studied, we obtained numerically that 
$\regul(h_{\gamma_i})<1$ and $\regul(h_{\gamma_i}^{-1})<1$, 
so \eqref{eq:holder-comp} yields that 
$H_{\gamma_1,\gamma_2}$ is not less regular than 
$\regul(h_{\gamma_1})\, \regul(h_{\gamma_2}^{-1})$.  
For $\gamma_1$ and $\gamma_2$ in the same universality class, 
however, we expect more 
-- because of ``cancellation'' of the ``singularities'' 
of $h_{\gamma_1}$ and $h_{\gamma_2}^{-1}$, 
we state our final 

\begin{conj}  \label{conj:inequalities}
The following inequalities hold for $i=1,2$:  
\[
\regul(h_{\gamma_i}) < \regul(H_{\gamma_1,\gamma_2}) \ , \qquad 
\regul(h_{\gamma_i}^{-1}) < \regul(H_{\gamma_1,\gamma_2}) \ .
\]
\end{conj}


\section{\label{sec:numerical-methods}Description of the numerical methods}

In this section we first describe the methods 
used for numerical computation 
of invariant circles 
and the related functions described 
in Sections \ref{sec:functions} 
and~\ref{sec:big-conj}.  
Then we briefly discuss the method 
we use to compute the global H\"older regularity 
of the functions.  


\subsection{\label{sec:computing-circles}
Computing critical invariant circles}

We need to compute (homotopically non-trivial) 
critical invariant circles of twist maps 
of the form \eqref{eq:standard}
with a Diophantine rotation number.  
We approximate such invariant circles by well-ordered periodic orbits 
(whose existence is guaranteed by 
the so-called Birkhoff's Geometric Theorem \cite{Birkhoff25}).  
Consider a sequence $\{\calX^{(j)}\}_{j\in\bbN}$ 
of well-ordered periodic orbits whose 
rotation numbers, $\{\rho_j\}_{j\in\bbN}$, 
constitute  a sequence of rational numbers 
which converge to a Diophantine number~$\rho$. 
Then the limit of these periodic orbits 
will be a well-ordered invariant set $\calX_{\rho}$ 
of rotation number~$\rho$; 
the existence of this set 
is guaranteed by Aubry-Mather theory \cite[Ch. 13]{KH}, 
\cite[Ch. 2]{Gole01}.  
The set $\calX_\rho$ can be a continuous curve 
which is a graph of a Lipschitz function 
under appropriate conditions 
(Theorem~\ref{th:mather}) 
or an orbit homeomorphic to a Cantor set 
(Cantorus).  

We approximate a Diophantine number $\rho$ 
by the rational numbers given by finite 
truncations of the continued fraction expansion of~$\rho$.  
In the case of the golden mean $\sigma_\mathrm{G}$ \eqref{eq:sigmaG}, 
these rational approximants are ratios $\rho_m=Q_{m-1}/Q_{m}$ 
of consecutive Fibonacci numbers~$Q_m$.  
The limit of the periodic orbits with rotation numbers 
$\rho_m$ is the invariant set $\calX_\rho$ 
we are looking for \cite{Greene79}.  

The problem of computing well-ordered orbits 
with a prescribed rational rotation number $\rho_m$ 
is greatly simplified if the function $V(\theta)$ 
in \eqref{eq:standard} is odd. 
In this case the task of finding a periodic orbit is reduced to
a one-dimensional problem 
because the map $F$ can be written 
as the composition of two involutions
as in \eqref{eq:Fdecomposed}; 
if such a decomposition is possible, 
the map $F$ is said to be {\em reversible}.  
If $F$ is reversible, there exists 
a set of straight lines in the $(\theta,r)$ space 
-- called {\em symmetry lines} -- 
that are invariant with respect to the maps $I_0$ and~$I_1$.  
It can be shown that any periodic orbit 
has two points that belong to one of these invariant 
straight lines, hence we can find these points 
(and, therefore, the periodic orbits that contain them) 
by using a one dimensional root finder \cite{Greene79}.  
Using the fact that the periodic orbits computed 
in this way are well-ordered, 
we can implement a numerical procedure 
to compute periodic orbits of several million points 
that approximate the invariant set~$\calX_\rho$.  

We are interested in studying the properties 
of area-preserving twist maps of the form \eqref{eq:standard}.  
When the parameter $\lambda$ in \eqref{eq:standard} 
is equal to 0, the corresponding twist map 
acts on each point $(\theta,r)$ 
as a rigid rotation in $\theta$-direction, 
$F(\theta,r) = (\theta+r,r)$, 
hence the phase space is foliated by invariant circles 
of the form $\{r=\mathrm{const}\}$.  
For small values of $|\lambda|$, KAM theory guarantees 
the existence of invariant circles with Diophantine rotation numbers.  
According to Conjecture~\ref{conj:cotas}, 
there is an upper bound $\Lambda(\rho)$ 
on the values of $|\lambda|$ 
such that for $|\lambda|<\Lambda(\rho)$ 
there exists an invariant circle with 
rotation number $\rho$; 
(some rigorous upper bounds on~$\Lambda(\rho)$ 
are given in Theorem~\ref{th:no-circles}).  
To find an accurate numerical approximation 
of the critical value, $\Lambda(\rho)$, of $\lambda$ 
for which the invariant circle of rotation number $\rho$ 
disintegrates, we applied an empirical method 
known as the ``residues method'' proposed in \cite{Greene79}, 
developed in \cite{OS87}, 
and partially justified rigorously in \cite{FalcoliniL92-Greene}.  
The main idea of this method is to determine 
the value of $\lambda$ such that the residue of all the 
approximating periodic orbits 
reaches the same value. Let $R_m$ be the residue of 
a periodic orbit which is the $m$th approximant 
to an invariant circle with rotation number~$\rho$.  
If $\lim_{m\to\infty} R_m = 0$, 
then there exists an invariant circle 
with rotation number~$\rho$; 
if $\lim_{m \to \infty} R_m = \infty$, 
then the invariant set $\calX_\rho$ becomes a Cantor set.  
A golden critical invariant circle is obtained 
at the value of $\lambda$ 
for which $R_m\simeq -2.55426$ for all values of~$m$.


\subsection{\label{sec:holder-numer}
Studying global H\"older regularity numerically}

In this section we describe briefly 
the method we employed to study global H\"older regularity, 
referring the reader to \cite{LlaveP02} 
for details, references, 
and assessment of the numerical accuracy 
of various numerical methods for computing regularity.  

In this paper, we will only use the method 
developed in \cite{LlaveP02} 
that was found to be the most accurate 
for studying global H\"older regularity 
-- the so-called "Continuous Littlewood-Paley" (CLP) method.  
Here we do not use the wavelet-based methods 
implemented in \cite{LlaveP02}.  
The CLP method has been used in~\cite{Carletti03,ApteLP05}.


\subsubsection{\label{sec:theo-CLP}Theoretical basis of the CLP method}

We recall the following definition.  

\begin{defi} 
For $\regul=n+\chi$ with $n\in \bbZ$,  $\chi\in
(0,1)$, we say that the function $K:\bbT\to\bbR$ 
has {\em (global) H\"older exponent} $\regul$ 
and write $K\in\Lambda_{\regul}(\bbT)$ when 
$K$ is $n$ times differentiable and, 
for some constant $C>0$, 
\[
|D^nK(\theta) - D^nK(\tilde{\theta})|  \leq C | \theta -
\widetilde{\theta}|^{\chi} 
\]
for all $\theta,\tilde\theta\in\bbT$.  
\end{defi}

For the case of an integer value of $\regul$, 
this definition is more complicated, 
but we will omit it since 
in the applications considered in this paper 
$\regul$ is not an integer.  

The following result can be found in 
\cite[Ch.~5, Lemma 5]{Stein70}.  

\begin{theo}[CLP] \label{th:clp}
 The function $K\in L^\infty(\bbT)$ 
is in $\Lambda_{\regul}(\bbT)$ if and
only if for some integer $\eta > \regul$ there exists a constant 
$C>0$ such that for any $t >0$
\begin{equation}  \label{diffB}
\left\| \left( \frac{\partial}{\partial t}\right)^{\eta} {\rm e}^{-t
\sqrt{-\Delta}} K \right\|_{L^{\infty}(\bbT)} \leq C t^{\regul-\eta} 
\, , 
\end{equation}
where $\Delta$ is the one-dimensional Laplace operator: 
$\Delta K(\theta) = K''(\theta)$.  
\end{theo}

\begin{rem}  
If the above result holds for some integer $\eta>\alpha$, then it
holds for all integers $\tilde\eta>\alpha$.  
\end{rem}

\begin{rem}
The operator ${\rm e}^{-t\sqrt{-\Delta}}$ 
is a convolution with the Poisson kernel: 
${\rm e}^{-t\sqrt{-\Delta}} K = P_{\exp(-2\pi t)} * K$.  
The function 
$u(\theta,t) := \mathrm{e}^{-t\sqrt{-\Delta}}K(\theta)$ 
is a solution of Laplace's equation, 
$u_{\theta\theta}+u_{tt} = 0$, 
on the half-cylinder $(\theta,t)\in \bbT\times (0,\infty)$, 
with Dirichlet boundary condition $u(\theta,0)=K(\theta)$.  
\end{rem}

\begin{rem} \label{rem:simil}
The mathematical theory only requires that 
\eqref{diffB} be an upper bound. 
In our numerical experiments, however, this bound is saturated 
for a significant range of values of~$t$.  
This fact is very possibly a consequence 
of the self-similarity at small scales 
of the functions we consider 
(which is at the basis of 
the renormalization group description).  
This saturation was also observed for the functions 
considered in \cite{LlaveP02,ApteLP05}.  
\end{rem}


\subsubsection{\label{sec:numer-implement}
Remarks on the numerical implementation}

To use the CLP method, we need to apply repeatedly 
Fast Fourier Transform (FFT), which is easiest to do 
if the values of the function $K$ in \eqref{diffB} 
are known at $2^N$ equally spaced points in the interval $[0,1)$ 
for some positive integer~$N$.  
However, as we describe in Sec.~\ref{sec:numerical-results}, 
we do not have control over the set of points 
at which the values of $K$ can be computed 
(where $K$ stands for any of 
the functions $R$, $g$, $h$, $h^{-1}$, $H$, $G$).  
Hence, the first step in applying the CLP method 
would be the computation of the values of $K$ on an evenly spaced grid.  
If we know accurately the values of $K$ at $M$ points in $[0,1)$, 
we can expect that by using some interpolation method, 
we will be able to obtain the approximate 
values of $K$ on $2^N\approx M$ equidistant points, 
$\{2^{-N}j\}_{j=0}^{2^N-1}$.  
To compute the approximate values of $K$ 
on the equidistant grid, we used cubic spline interpolation.  
Using interpolation poses the question 
of whether the interpolated values 
represent faithfully the true values of~$K$.  
Naturally, the answer to this question is no, 
but practically if $M$ is large enough, 
the interpolated values will be very close to the true values, 
which will allow us to compute many Fourier coefficients 
of $K$ accurately.  
The degree of ``contamination'' of the Fourier spectra 
due to the interpolation depends on the uniformity of the distribution of the 
$M$ points at which the values of $K$ 
is known accurately (see Remark~\ref{rem:noisy-spec-hm1}).  

To apply the CLP method numerically, 
we observe that the operator 
$\left(\frac{\partial}{\partial t}\right)^{\eta} 
{\rm e}^{-t\sqrt{-\Delta}}$ 
used in Theorem~\ref{th:clp} 
is diagonal in a Fourier series representation: if 
$K(\theta) = \sum_{k\in\bbZ} \hat{K}_k {\rm e}^{-2\pi i k \theta}$, 
then
\begin{equation}  \label{diffG}
\left(\frac{\partial}{\partial t}\right)^{\eta} 
{\rm e}^{-t\sqrt{-\Delta}} K(\theta) 
= 
\sum_{k\in\bbZ} (-2 \pi |k|)^{\eta}\,
{\rm e}^{-2\pi t |k| } \, \hat{K}_k 
\, {\rm e}^{-2\pi i k \theta}
\, .
\end{equation}
Having computed the values of the spline interpolant 
to the function $K$ on an equally spaced grid, 
applying \eqref{diffB} is easy.  
Namely, we fix some values of the parameters $\eta$ and $t$, 
perform FFT to find 
$\hat K_k$ and compute the Fourier coefficients 
of 
$\left(\frac{\partial}{\partial t}\right)^{\eta} 
{\rm e}^{-t\sqrt{-\Delta}} K$.  
Then we apply inverse FFT to find the values of 
$\left(\frac{\partial}{\partial t}\right)^{\eta} 
{\rm e}^{-t\sqrt{-\Delta}} K$ 
at the equally spaced set of points 
$\{2^{-N}j\}_{j=0}^{2^N-1}$; 
among these values we find the one 
with maximum absolute value 
-- this value we take for the numerical value of 
the left-hand side of~\eqref{diffB}.  
For a fixed value of $\eta$, 
we repeat this procedure for many values of~$t$ 
(we used several hundred values of $t$ in our computations).  
According to \eqref{diffB}, if we plot 
\begin{equation}  \label{eq:loglog}
\log \left\|\left(\frac{\partial}{\partial t}\right)^{\eta}\,
{\rm e}^{-t\sqrt{-\Delta}} K \right\|_{L^{\infty}(\bbT)} 
\quad \mbox{versus} \quad 
\log t \ , 
\end{equation}
the points should lie below a straight line of slope $\regul-\eta$.  
As pointed out in Remark~\ref{rem:simil} 
(see also Remark~\ref{rem:fluctuations}) 
the points on the log-log plot should not only be below 
this straight line, but should close to it.  
We perform linear regression to find the slope of this line, 
from which we find~$\regul$.  



\section{Numerical results}   \label{sec:numerical-results}

\subsection{\label{sec:maps-studied}Twist maps studied}

We study numerically a set of one-parameter families of area-preserving 
twist maps of the form~\eqref{eq:standard}, 
each family having a different function~$V$.  
Within each family we find numerically 
the value $\Lambda(\sigma_{\mathrm{G}})$ 
of the parameter $\lambda$ 
for which the golden invariant circle is critical.  
The set of functions $V$ (all of them odd) 
that we selected is the following:  
\begin{enumerate}
  \item The standard (Taylor-Chirikov) map: 
\begin{equation}  \label{eq:armo_1}
V_1(\theta)= -\frac{1}{2 \pi} \sin 2 \pi \theta \ .
\end{equation}

  \item The ``standard map with two harmonics'': 
\begin{equation}  \label{eq:armo_101}
V_2(\theta)= -\frac{1}{2 \pi}
\left[ \sin(2 \pi \theta) - 0.03 \sin(6 \pi \theta) \right] \ .
\end{equation}

  \item The ``critical standard map with two harmonics'': 
\begin{equation}  \label{eq:two}
V_3(\theta)= -\frac{1}{2 \pi} 
\left[ \sin(2 \pi \theta) - \frac{1}{2} \sin(6 \pi \theta) \right] \ .
\end{equation}
For this choice of coefficients, 
the first three derivatives of $V(\theta)$ at $\theta=0$ are zero.  

 \item The ``$0.2$-analytic map'': 
\begin{equation}  \label{eq:ana_0.2}
V_4(\theta)= -\frac{1}{2 \pi} \frac{\sin(2 \pi \theta) }
{ 1-0.2\cos(2 \pi \theta) } \ .
\end{equation} 
This map has infinitely many nonzero Fourier coefficients.  
It would be very interesting to study 
this map when the coefficient of the cosine function 
in the denominator is close to~1, 
but then it would be extremely difficult to compute 
periodic orbits.  

 \item The ``$0.4$-analytic map'': 
\begin{equation}  \label{eq:ana_0.4}
V_5(\theta)= -\frac{1}{2 \pi} \frac{\sin(2 \pi \theta) }
{ 1-0.4\cos(2 \pi \theta) } \ .
\end{equation} 

 \item The ``tent map'': 
\begin{equation}  \label{eq:trian}
V_6(\theta)= \sum_{j=1}^{17} c_j \sin(2 \pi j \theta) \ ,
\end{equation} 
where 
$c_j = (-1)^{\frac{j+1}{2}} \frac{4}{\pi^2 j^2}$ 
for $j$ odd, and 
$c_j = 0$ for $j$ even, 
are the Fourier coefficients 
of the function 
\[
\calV(\theta) 
= 
\left\{
\begin{array}{ll}
	-4\theta \ , & \mbox{ for } 0 \leq \theta < \frac14 \ , \\
	4\theta - 2 \ , & \mbox{ for } \frac14 \leq \theta < \frac34 \ , \\
	4 - 4\theta \ , & \mbox{ for } \frac34 \leq \theta < 1 \ .
\end{array}
\right. 
\]
The function $V_6$ is close to 
the piecewise linear continuous function $\calV$.  

\end{enumerate}

   Our numerical experiments were performed with 
the twist maps coming from the above six functions~$V(\theta)$ 
and the corresponding values $\Lambda(\sigma_\mathrm{G})$, 
following the steps below.  

\begin{enumerate}

\item 
As discussed in Sec.~\ref{sec:computing-circles}, 
the invariant circle of rotation number 
$\sigma_{\mathrm{G}}$ can be obtained as a limit 
of periodic orbits 
of rotation numbers equal to ratios of consecutive 
Fibonacci numbers, $\rho_m=Q_{m-1}/Q_{m}$.  
We chose to compute hyperbolic periodic orbits, 
and found the values of 
$\Lambda(\sigma_\mathrm{G})$ 
by applying Greene's residue criterion.  

\item 
The highest approximant to the critical invariant circle 
that we computed was a periodic orbit 
with rotation number $Q_{29}/Q_{30}= 832040/1346269$.  
The value of $\Lambda(\sigma_{\mathrm{G}})$ was determined 
by using the condition that the difference, 
$|R_{30}-R_{29}|$ of the residues of the 
periodic orbits with periods $Q_{29}$ and $Q_{30}$ 
be zero 
(in practice, we wanted that this difference 
be smaller than~$10^{-10}$).  
The periodic orbits were computed with an
error not exceeding~$10^{-23}$.

\item 
We computed the hyperbolic periodic orbit 
$\left\{(\theta_m,r_m)\right\}_{m=0}^{M-1}$ 
of period $M=Q_{30}$.  
The values of the advance map $g$ \eqref{eq:g-def} 
at the points $\theta_m$ ($m=0,1,\ldots,M-1$) 
were then computed by 
$g(\theta_m) = \theta_{m+1}$ 
(here and below, 
we take $\mathrm{mod}\,1$ wherever needed).  
The values of the conjugacy $h$ 
at the points $m\sigma_\mathrm{G}$ 
(which corresponds to $m$ applications 
of the rigid rotation by $\sigma_\mathrm{G}$ to~0) 
are given by 
$h(m\sigma_\mathrm{G})=\theta_m$, 
and, similarly, 
$h^{-1}(\theta_m)= m\sigma_\mathrm{G}$.  

\item
In our Fourier-analysis-based CLP method 
we need to deal with periodic functions, 
so we compute the ``periodized'' versions, 
$g-\mathrm{Id}$, 
$h-\mathrm{Id}$, 
$h^{-1}-\mathrm{Id}$, 
of the functions $g$, $h$, and~$h^{-1}$.  
Then we sort the periodized functions 
with respect to their argument; 
the function $R$ is already periodic, 
so we just sort its values.  

\item 
The periodic functions are 
passed to the cubic spline interpolation routine 
to find approximations to the values 
of the corresponding functions 
on a uniformly spaced grid of $2^N$ points; 
we used $N=20$.  

\item
The interpolated values of the functions 
are given to the CLP algorithm 
to compute their H\"older regularity.  
We used integer values of $\eta$ in \eqref{diffB} 
from 1 to 5, and for each analyzed function 
chose the value of $\eta$ 
that gave the best straight line 
on the log-log plot~\eqref{eq:loglog}.  
The log-log plots for the other values of $\eta$ 
were used as a consistency check.  

\end{enumerate}

\begin{rem}  \label{rem:compute-H}
In computing the big conjugacies $H_{\gamma_1,\gamma_2}$, 
we had to take special care of the preserving the symmetries 
of the maps $h$.  
For each critical circle we studied, 
we needed to find the appropriate value 
of the constant $\zeta$ and shift the argument 
of the corresponding function $h$ 
as explained in Sec.~\ref{sec:H-sym}.  
\end{rem}


\subsection{\label{sec:visual}
Critical invariant circles -- visual explorations}

In Fig.~\ref{fig:R} we show the critical invariant circles 
which, by the definition \eqref{eq:R-def}, 
are graphs of the functions $R$ 
corresponding to the six twist maps studied.  
%
%
The graphs of the ``periodized versions'' 
of the advanced maps, $g-\mathrm{Id}$, 
the conjugacies, $h-\mathrm{Id}$, 
and their inverses, $h^{-1}-\mathrm{Id}$, 
are plotted in Figures \ref{fig:g}, 
\ref{fig:h-shifted}, and~\ref{fig:all-hm1}, resp.  

%
%

Figure~\ref{fig:zoom} illustrates the self-similar nature 
the functions $h$; 
needless to say, the insets are true zooms of parts 
of the graph of the function.  
%
%

Fig.~\ref{fig:bigH} shows the graphs 
of several periodized big conjugacies $H-\mathrm{Id}$; 
it is obvious that these functions 
are smoother than the ``small'' conjugacies~$h$.  
%
%



\subsection{\label{sec:spectra}Fourier spectra, CLP method}

Fig.~\ref{fig:spectrum-h} depicts $\log_{10}$ of the 
modulus of the $k$th Fourier coefficient 
of a periodized conjugacy $(h-\mathrm{Id})$ 
versus~$\log_{10} k$; here $h$ is the conjugacy 
corresponding to the twist map $F$ with~$V_3$~\eqref{eq:two}.  
The horizontal distance between two adjacent high peaks 
is approximately equal to $|\log_{10} \golden|\approx 0.209$, 
which is a manifestation of the self-similarity at small scales.  
%
%
The $\log_{10}$-$\log_{10}$ plots of the 
Fourier spectra of the functions $(g-\mathrm{Id})$ 
and $(h^{-1}-\mathrm{Id})$ for the same map~$F$ 
are given in Fig.~\ref{fig:spectrum-two-g-hm1}.  
%
%

\begin{rem}  \label{rem:noisy-spec-hm1}
Note that the spectrum of $h$ 
is very accurate 
even at length-scales $\sim 10^{-6}$, 
while the spectrum of $h^{-1}$ 
is quite noisy.  
As explained in Sections \ref{sec:numer-implement} 
and~\ref{sec:maps-studied}, 
the main reason for this is that 
the exact values of $h$ are known at the points 
$(m\sigma_\mathrm{G})\,\mathrm{mod}\,1$, 
which are almost uniformly distributed on~$\bbT$.  
On the other hand, we know the exact values 
of of $g$ and $h^{-1}$ at the very nonuniformly 
distributed points of the form $g^m(\theta_0)$ 
(because the underlying invariant measure 
is singular -- see Sec.~\ref{sec:sing-measures}), 
which results in the presence 
of big gaps between these points 
and, hence, distorted values of the spline interpolant.  
\end{rem}


In Fig.~\ref{fig:clp-slopes} 
%
%
we show several plots of 
$\log_{10}$ of the left-hand side of \eqref{diffB} 
versus $\log_{10}t$.  
The six lines in each group 
of lines of similar slope correspond to the six 
different choices \eqref{eq:armo_1}--\eqref{eq:trian} 
of functions~$V$, 
and the lines in each group come 
from the same value of~$\eta$ in~\eqref{diffB}.  
Each on the ``lines'' in the figure in fact 
consists of 400 points (visually indistinguishable).  
The computer time spent on the CLP analysis 
is of the order of one minute per point 
(we used $2^{20}$ Fourier coefficients to compute each of these points).  

We computed the regularity 
by performing linear regression on the points 
on graphs like the one in Fig.~\ref{fig:clp-slopes}, 
in the regions where the points 
follow more or less a straight line.  
As one can see from this figure, 
for $t$ close to~$1$ 
(i.e., $\log_{10}t\approx 0$), 
the graphs for different functions 
are not straight lines, then as $t$ decreases 
they form more or less straight lines, 
and as $t$ decreases further, these lines level out.  
This behavior 
can be understood intuitively from \eqref{diffG} 
-- for $t\approx 1$ 
the high-$k$ Fourier coefficients 
are strongly suppressed by the factor 
$(-2\pi|k|)^\eta\,{\rm e}^{-2\pi t|k|}$, 
so the CLP method still does not ``feel'' 
the asymptotic self-similarity 
of the functions at small length-scales; 
in the other extreme, 
the leveling-out of the lines 
for very small $t$ 
comes from the fact that in our computations 
we use a finite -- albeit very large -- 
number of Fourier coefficients.  

\begin{rem}   \label{rem:fluctuations}
The ``straight lines'' in Fig.~\ref{fig:clp-slopes} 
%
%
are not really straight 
(which has been noticed 
in different contexts in~\cite{LlaveP02,ApteLP05}).  
We show this effect in Fig.~\ref{fig:clp-diffs}, 
which was created as follows.  
We took the six lines for $\eta=2$ 
from Fig.~\ref{fig:clp-slopes}, 
and for each of them we computed 
the slope of the line as a function 
of the horizontal coordinate in the figure, $\log_{10}t$.  
To compute this slope, we took each pair 
of adjacent points on the line and found the slope 
of the straight line connecting these points.  
The distance between two consecutive peaks 
in Fig.~\ref{fig:clp-slopes} is $|\log_{10}\golden|$; 
more interestingly, as $\log_{10}t$ becomes more negative, 
the lines tend to the same wavy line, 
until all lines reach saturation 
around $\log_{10}t\approx -4.5$.  
\end{rem}


\subsection{Global H\"older regularities 
-- numerical results}   \label{sec:reg-results}

Table~\ref{table:regul-ghR} 
%
%
summarizes our numerical results.  
The first column 
gives the map $V$ used 
in the numerical computations 
(for the six functions $V$ given 
by \eqref{eq:armo_1}--\eqref{eq:trian}).  
In the other columns we give the the values of the 
(global) H\"older exponent $\regul$ of the function $R$ 
(representing the invariant circle as a graph 
in the $(\theta,r)$-plane), 
the advance map $g$, the conjugacy $h$ and its inverse,~$h^{-1}$, 
coming from the (dynamics on) 
the golden critical invariant circle 
of the corresponding area-preserving twist map~$F$.  
The notations used are the following: 
$1.85(15)$ stands for $1.85\pm0.15$, 
and $0.726(3)$ for $0.726\pm0.003$.  
Note that within the numerical error, 
$\regul(R)=\regul(g)$, as expected.  

We also computed the H\"older regularities 
of all big conjugacies~$H$ 
between each of the six functions $h_1$, $\ldots$, $h_6$ 
(coming from $V_1$, $\ldots$, $V_6$) 
with all other $h_j$'s.  
We applied the CLP method to find that 
the regularity of all thirty functions $H$ studied is 
\begin{equation}  \label{eq:regul-H}
\regul(H) = 1.80 \pm 0.15 \ .
\end{equation}



\section{\label{sec:discussion}Discussion and conclusion}

In Sections~\ref{sec:scaling} and \ref{sec:sing-measures} 
we point out some relationships between 
our results and previous studies 
related to universal scaling factors 
and singular measures.  
In the final Section~\ref{sec:conclusion}, 
we recapitulate our findings.


\subsection{\label{sec:scaling}H\"older regularity and scaling factors}

Here we will explain how the scaling 
of the distances of closest returns 
of the iterates of a point 
gives bounds on the H\"older regularity 
of some of the functions we study.  
Our analysis here is reminiscent 
of the analysis in \cite[Sec.~8.2]{LlaveP02}.  

We start by recalling the crucial observation 
of Kadanoff and Shenker \cite{Kadanoff81,ShenkerK82} 
(see also \cite[Sec.~4.4]{MacKay82}) 
of the existence of universal scalings 
in the distribution of the iterates of 
the Taylor-Chirikov map on the critical invariant circle~$\gamma$ 
in neighborhoods of certain points of~$\gamma$.  
Let $\rarefied\in\bbT$ stand for the value 
around which the iterates of the function $g$ 
are most rarefied (in our notations $\rarefied=\frac12$, 
while in \cite{ShenkerK82} $\rarefied=0$).  
Let $\dense\in\bbT$ stand for the value 
around which the iterates of the function $g$ 
are most dense (in our notations $\rarefied=0$, 
while in \cite{ShenkerK82} it is $\rarefied=\frac12$).  
Since by Theorem~\ref{th:mather} the function $R$ is Lipschitz, 
around the points 
$(\rarefied,R(\rarefied))$ 
and 
$(\dense,R(\dense))$, 
the iterates of any point on $\gamma$ under $F$ 
are most rarefied, resp.\ dense.  
Shenker and Kadanoff found that 
the critical invariant circle 
in a neighborhood of $\rarefied$ 
is asymptotically invariant under 
simultaneous scalings in both $\theta$- and $r$-directions, 
with scaling factors 
\[
\alpha_0 \approx -1.414836 \ \ \mbox{(in $\theta$)} \ , 
\qquad 
\beta_0 \approx -3.0668882 \ \ \mbox{(in $r$)} 
\]
(see also the bounds on these values 
in Stirnemann \cite{Stirnemann97}).  
This implies that, for large $n$, 
\begin{equation}  \label{eq:scaling-step1}
\frac{g^{Q_{n+1}}(\rarefied)-\rarefied}{g^{Q_n}(\rarefied)-\rarefied}
\approx
\alpha_0^{-1} \ ,
\qquad 
\frac{R(g^{Q_{n+1}}(\rarefied))-R(\rarefied)}
		{R(g^{Q_n}(\rarefied))-R(\rarefied)} 
\approx
\beta_0^{-1} \ .
\end{equation}
The scaling around $\dense$ is a bit more 
complicated -- it is called ``step-3'' scaling 
for obvious reasons:
\begin{equation}  \label{eq:scaling-step3}
\frac{g^{Q_{n+3}}(\dense)-\dense}{g^{Q_n}(\dense)-\dense}
\approx
\alpha_3^{-1} \ ,
\qquad 
\frac{R(g^{Q_{n+3}}(\dense))-R(\dense)}
		{R(g^{Q_n}(\dense))-R(\dense)} 
\approx
\beta_3^{-1} \ ,
\end{equation}
where the ``step-3'' scaling factors are 
\[
\alpha_3 \approx -4.84581 \ \ \mbox{(in $\theta$)} \ , 
\qquad 
\beta_3 \approx -16.8597 \ \ \mbox{(in $r$)} \ .
\]

To understand heuristically why these scalings 
give restrictions on the H\"older regularity of $R$, 
set 
$\Delta \theta := g^{Q_{n+1}}(\rarefied)-\rarefied$, 
$\Delta r := R(g^{Q_{n+1}}(\rarefied))-R(\rarefied)$ 
for some large value of~$n$.  
Then if the local H\"older exponent of $R$ 
at $\theta=\theta_{\mathrm{rar}}$ is $\regul$, 
we will have 
$|\Delta r| \sim |\Delta \theta|^\regul$.  
If the graph of $R$ is asymptotically invariant 
around $(\rarefied,R(\rarefied))$ 
with respect to the scalings \eqref{eq:scaling-step1}, 
we will have 
$|\beta_0 \, \Delta r| \sim |\alpha_0 \, \Delta \theta|^\regul$.  
``Dividing out'' the last two relationships, 
we obtain $|\beta_0| \sim |\alpha_0|^\regul$, 
i.e., 
$\regul\sim\frac{\log|\beta_0|}{\log{|\alpha_0|}}$.  
This argument (which can easily be made rigorous) 
implies that the (global) H\"older exponent of $R$ 
does not exceed 
$\frac{\log|\beta_0|}{\log{|\alpha_0|}} \approx 3.22945$.  
The scaling \eqref{eq:scaling-step3} 
yields a tighter bound on the H\"older regularity 
of~$R$: 
\begin{equation}  \label{eq:reg-R-bound}
\regul(R) 
\leq 
\frac{\log|\beta_3|}{\log{|\alpha_3|}} 
\approx 
1.7901 \ .
\end{equation}  
Note that the fact that the scaling \eqref{eq:scaling-step3} 
is ``step-3'' (as opposed to ``step-1'') 
is irrelevant for the bounds on the H\"older regularity.  

To obtain bounds on $\regul(h)$ and $\regul(h^{-1})$, 
we use Lemma~8.1 from \cite{LlaveP02}, 
which says that if the function $h$ conjugates $f_1$ and $f_2$, 
$h\circ f_1=f_2\circ h$, and if for some sequence 
of positive integers $Q_n$ the functions 
$f_j$ ($j=1,2$) behave in a neighborhood 
of the fixed point 
$\theta_\mathrm{fix}=h(\theta_\mathrm{fix})$ 
of $h$ as follows:
\[
f_j^{Q_n}(\theta_\mathrm{fix}) 
= 
\theta_\mathrm{fix} + C_j \eta_j^{-n} + o(\eta_j^{-n}) 
\]
for some constants $\eta_j$ and $C_j$, 
then 
$\regul(h) \leq \frac{\log|\eta_2|}{\log|\eta_1|}$.  
Applying this 
to the definition of $h$ and using the well-known fact 
that $(Q_n\,\golden)\,\mathrm{mod}\,1\leq C\sigma_\mathrm{G}^n$, 
we obtain the bounds 
\begin{equation}  \label{eq:bound-regul-h-hm1}
\regul(h) 
\leq 
\frac{\log|\alpha_0^{-1}|}{\log|\golden|}  
\approx  
0.721125 \ , \qquad 
\regul(h^{-1}) 
\leq 
\frac{\log|\sigma_\mathrm{G}^3|}{\log|\alpha_3^{-1}|} 
\approx 
0.91478 \ .
\end{equation}
A comparison with Table~\ref{table:regul-ghR} suggests that these bounds are saturated.  


\subsection{\label{sec:sing-measures}Conjugacies and singular measures}

The functions whose H\"older regularity we study 
are defined through high iterates of maps.  
For example, the graph of the function $R$ defined by \eqref{eq:R-def} 
is nothing but the the critical invariant circle $\gamma$ of~$F$ 
which is is filled densely by the iterates 
$F^n(\theta_0,r_0)$ of some point $(\theta_0,r_0)\in\gamma$.  
Here we discuss how 
some characterizations of the singularities 
in the distribution of the iterates 
of $F$ on~$\gamma$ 
are related to the H\"older regularity 
of some of the functions considered.  

Hentschel and Procaccia \cite{HentschelP83} 
pointed out the importance of 
the {\em generalized (R\'enyi) dimensions} $D(q)$ 
of a singular measure for dynamical systems;  
these quantities have been defined previously 
in the context of probability theory 
by R\'enyi~\cite{Renyi60}.  
Halsey et al in their seminal paper \cite{HalseyJKPS86} 
related heuristically the R\'enyi dimension of a singular measure 
to the {\em spectrum of singularities~$f(\alpha)$}.  
We recall that $f(\alpha)$ is the Hausdorff dimension of the set $E_\alpha$ 
of points where the measure has singularity of strength~$\alpha$.  
The spectrum $f(\alpha)$ is a function supported on the interval 
$[\alpha_\mathrm{min},\alpha_\mathrm{max}]$, 
where 
$\alpha_\mathrm{min} = D(\infty)$, 
resp.\ $\alpha_\mathrm{max} = D(-\infty)$, 
describe the scaling behavior 
of the measure in the region where the measure 
is most dense, resp.\ most rarefied.  

Let $(\theta_0,r_0)$ be an arbitrary point 
on the critical invariant circle~$\gamma$ of the 
area-preserving twist map~$F$.  
Then the distribution of the iterates in a very long orbit, 
$\{F^n(\theta_0,r_0)\}_{n=0}^K$, approaches as $K\to\infty$ 
the ``density'' of the measure on $\gamma$ 
that is invariant with respect of the restriction 
of the map $F$ onto~$\gamma$.  
(We put ``density'' is quotation marks 
because for singular measures this is not 
a function, but a set of Dirac $\delta$-distributions.)  
This invariant measure on $\gamma$ induces an invariant 
measure $\mu_g$ of the map $g$ on~$\bbT$.  
It is easy to see that \eqref{eq:h-def} 
implies that 
\[
h^{-1}(\theta) = \int_0^\theta d \mu_g 
\]
(for an appropriately chosen $\zeta$ in the redefinition 
of $h$ as in Sec.~\ref{sec:H-sym}).  
This relationship implies that the spectrum 
of singularities $f(\alpha)$ of the measure $\mu_g$ 
is the same as the {\em H\"older spectrum} 
$f_\mathrm{H}(\alpha)$ of the function~$h^{-1}$.  
By definition, $f_\mathrm{H}(\alpha)$ 
is the Hausdorff dimension of the set where 
the local H\"older exponent of the function 
is equal to~$\alpha$; 
for a readable account we refer the reader 
to Jaffard~\cite{Jaffard97}.  
The (global) H\"older regularity $\regul(\phi)$ 
of a function $\phi$ 
is equal to the lowest end, $\alpha_\mathrm{min}$, 
of the support of the H\"older spectrum, 
$f_\mathrm{H}(\alpha)$, of~$\phi$.  

Osbaldestin and Sarkis \cite{OsbaldestinS87} 
applied the method of \cite{HalseyJKPS86} 
to determine numerically the functions $f(\alpha)$ 
and $D(q)$ of the invariant measure~$\mu_g$ 
coming from the distribution of iterates 
of the Taylor-Chirikov map $F$ 
on the golden invariant circle.  
They found that 
\[
\alpha_\mathrm{min} = D(\infty) \approx 0.915 \ , 
\qquad 
\alpha_\mathrm{max} = D(-\infty) 
	\approx 1.387 \approx \frac{1}{0.720} \ .
\]
Comparing with the values in Table~\ref{table:regul-ghR}, 
the reader should recognize that their $\alpha_\mathrm{min}$ 
is nothing but our $\regul(h^{-1})$, 
while $\alpha_\mathrm{min}$ is equal to the inverse 
of the regularity of the conjugacy~$h$.  

Buri\'c et al \cite{BuricMT97,BuricMT98} 
studied numerically 
the Taylor-Chirikov map and 
the map \eqref{eq:standard} with 
$V(\theta) = \frac12 \sin 2\pi \theta + \frac14 \sin 4 \pi \theta$, 
for rotation numbers with 
continued fraction expansions of the form 
$[S,1^\infty]:=[S,1,1,1,\ldots]$, 
$[S,2^\infty]$, $[S,3^\infty]$, $[S,4^\infty]$, 
where $S$ stands for some short string of positive integers.  
They found that $f(\alpha)$ and $D(q)$ 
depend only on the tail 
but do not depend on the initial part~$S$ 
as well as on whether the Taylor-Chirikov map 
or the other map was used in their numerics.  

Other papers related to numerical computations 
of singular measures 
on critical invariant circles of area-preserving 
twist maps are Shi and Hu \cite{ShiH91,HuS94}, 
where the methods of \cite{HalseyJKPS86} were used, 
and Hunt et al \cite{HuntKSY96}, 
where the authors used the thermodynamic formalism 
developed in \cite{VulSK84} 
to compute the information dimension $D(1)$ 
of the standard map for different rotation numbers.


\subsection{\label{sec:conclusion}Conclusion}

We computed accurately the golden critical invariant circles 
for six twist maps of the form \eqref{eq:standard} 
and the global H\"older regularity $\kappa$ 
of some functions related to the dynamics on these circles.  
Our numerical experiments lend credibility 
to Conjectures \ref{conj:regularity}, \ref{conj:stratified} 
and \ref{conj:big-conj} 
concerning the universality of the regularities 
of the functions $R$, $g$, $h$, $h^{-1}$ and~$H$ 
(see Table~\ref{table:regul-ghR} and~\eqref{eq:regul-H}).  
Yamaguchi and Tanikawa \cite{YamaguchiT99} 
found numerically that the golden invariant circle 
(given by the function $R$) of the Taylor-Chirikov map is differentiable 
but $R'$ is not of bounded variation; 
our studies significantly narrow 
the numerical bounds on~$\regul(R)$ for this and for other maps.   

Our results seem to indicate 
that the regularities of $R$, $h$, and $h^{-1}$ 
saturate the upper bounds 
\eqref{eq:reg-R-bound} and \eqref{eq:bound-regul-h-hm1} 
coming from previous studies of scaling exponents.  

Our finding that $\regul(H)$ is greater 
than $\regul(h)$ and $\regul(h^{-1})$ 
by a comfortable margin 
(cf.\ Conjecture~\ref{conj:inequalities}) 
has an interesting consequence.  
As discussed in Sec.~\ref{sec:sing-measures}, 
the H\"older regularity of $h$ and $h^{-1}$ 
is different at different points, 
and for each 
$\alpha\in(\alpha_\mathrm{min}, \alpha_\mathrm{max})$, 
the set $E_\alpha$ 
(where the pointwise H\"older exponent of~$h^{-1}$ is~$\alpha$) 
has Hausdorff dimension $f_H(\alpha)$ 
strictly between $0$ and~$1$.  
Previous numerical studies indicated that $f_H(\alpha)$ 
are the same for different maps~$F$.  
Our finding shows that the ``irregularities'' 
of functions $h$ coming from different maps $F$ 
are interspersed in the same way in $[0,1]$ 
for all twist maps studied.  
Note that this does not mean 
that for a certain value of $\alpha$ 
the sets $E_\alpha$ are {\em the same} 
for different $F$ in the same universality class 
-- only the way all sets $E_\alpha$ 
for different $\alpha$ are interwoven is universal.  

%

It would be interesting to apply wavelet-maxima methods 
for pointwise regularity \cite{ArneodoBM95,MuzyBA94} 
(see also the rigorous analysis in \cite{Jaffard97}) 
to the problem studied in this paper 
and to compare the results of the wavelet analysis 
with the results about the singular invariant measures.  

As a by-product of our studies, 
we have computed millions of Fourier coefficients 
of the functions~$h$, 
and noticed some self-similarity properties 
that to the best of our knowledge 
have not been observed before.  
Presently we are working on understanding these properties.  


%


\pagebreak

\section*{Acknowledgments}

We would like to express our gratitude 
to Rafael de la Llave, 
who introduced the authors 
of the present paper to each other, 
suggested the problem, 
and took an active part in the early stages 
of this research.  We have profited immensely 
from his expert advice and friendly prodding 
throughout our work on the paper.  

The research of NP was partially supported 
by National Science Foundation grant DMS-0405903 
and by the Michigan Center for Theoretical Physics 
(where part of this research was conducted).  
The authors would like to thank 
IIMAS-UNAM for supporting NP's visits to Mexico City.  

Our computations were carried out 
on the computers of IIMAS-UNAM 
and the Department of Mathematics 
of the University of Texas.  
AO would like to thank Ana P\'erez for the computational support.  
We used the doubledouble software developed by 
Keith Briggs, 
and the convenient plotting tool Grace \cite{Grace} 
(a descendant of ACE/gr developed by Paul J.\ Turner).  
We express our thanks to 
all these people and organizations.



\pagebreak

\bibliographystyle{plain}  

\begin{thebibliography}{88} 



\bibitem
{Kadanoff81}
\Author{L. Kadanoff}
\ArticleTitle{Scaling for a critical Kolmogorov-Arnold-Moser 
	trajectory}
\JournalTitle{Phys. Rev. Lett.} {\bf 47} 
(1981), 1641--1643.
\bibitem
{ShenkerK82}
\Author{S. Shenker and L.P. Kadanoff}
\ArticleTitle{Critical behavior on a KAM surface:  
I. Empirical results}  
\JournalTitle{J. Stat. Phys.} {\bf 27} 
(1982), 631--656.  
\bibitem
{HalseyJKPS86}
\Author{T. C. Halsey, M. H. Jensen, L. P. Kadanoff, I. Procaccia, 
	B. I. Shraiman}
\ArticleTitle{Fractal measures and their singularities: 
	The characterization of strange sets}
\JournalTitle{Phys. Rev. A} {\bf 33} 
(1986), 1141--1151, 
reprinted in \cite[pp.\ 540--550]{Cvitanovic89}.  
\bibitem
{LlaveP02}
\Author{R. de la Llave and N. Petrov}
\ArticleTitle{Regularity of conjugacies between critical circles maps:
An experimental study}
\JournalTitle{Experiment. Math.} {\bf 11} (2002), 219--241.
\bibitem
{TresserC78}
\Author{C. Tresser and P. Coullet}
\ArticleTitle{It\'erations d'endomorphismes et groupe de renormalisation}
\JournalTitle{C. R. Acad. Sci. Paris S\'er. A-B} {\bf 287} (1978), A577--A580.
\bibitem
{Feigenbaum78}
\Author{M. J. Feigenbaum}
\ArticleTitle{Quantitative universality for a class 
	of nonlinear transformations}
\JournalTitle{J. Stat. Phys.} {\bf 19} (1978), 25--52.
\bibitem
{Feigenbaum79}
\Author{M. J. Feigenbaum}
\ArticleTitle{The universal metric properties 
	of nonlinear transformations}
\JournalTitle{J. Stat. Phys.} {\bf 21} (1979), 669--706, 
reprinted in \cite[pp.\ 207--244]{Cvitanovic89}.  
\bibitem
{Shenker82}
\Author{S. Shenker}
\ArticleTitle{Scaling behavior in a map of a circle onto itself: 
  empirical results}
\JournalTitle{Phys. D} {\bf 5} 
(1982), 405--411, 
reprinted in \cite[pp.\ 405--411]{Cvitanovic89}.  
\bibitem
{ColletEL80}
\Author{P. Collet, J. P. Eckmann and O. E. Lanford, III}
\ArticleTitle{Universal properties of maps on an interval}
\JournalTitle{Comm. Math. Phys.} {\bf 76} (1981), 211--254.
\bibitem
{FKS82}
\Author{M. J. Feigenbaum, L. P. Kadanoff, and S. J. Shenker}
\ArticleTitle{Quasiperiodicity in dissipative systems: 
  a renormalization group analysis}
\JournalTitle{Phys. D} {\bf 5} 
(1982), 370--386.
\bibitem
{ORSS83}
\Author{S. Ostlund, D. Rand, J. Sethna, and E. Siggia}
\ArticleTitle{Universal properties of the transition 
  from quasiperiodicity to chaos in dissipative systems}
\JournalTitle{Phys. D} {\bf 8} 
(1983), 303--342.
\bibitem
{MacKay83}
\Author{R. S. MacKay}
\ArticleTitle{A renormalisation approach to invariant circles in
area preserving twist maps}
\JournalTitle{Phys. D} {\bf 7} (1983), 283--300, 
reprinted in \cite[pp.\ 462--479]{MacKayM87}.  
\bibitem
{Meiss92}
\Author{ J. D. Meiss}
\ArticleTitle{Symplectic maps, variational principles, and transport}
\JournalTitle{Rev. Modern Phys.} {\bf 64} (1992), 795--848.
\bibitem
{Gole01}
\Author{C. Gol\'e}
\BookTitle{Symplectic Twist Maps}
World Scientific, 2001.
\bibitem
{MacKay82}
\Author{R. S. MacKay}
\ArticleTitle{Renormalization in Area Preserving Maps}
\BookTitle{PhD Thesis}
Princeton University, 1982; 
published with notes as R. S. MacKay, 
{\em Renormalization in Area-Preserving Maps}, 
World Scientific, Singapore, 1993.  
\bibitem
{Birkhoff20}
\Author{G. D. Birkhoff}
\ArticleTitle{Surface transformations and their 
	dynamical applications}  
\JournalTitle{Acta Math.} {\bf 43} (1920), 1--119, 
reprinted in \cite[pp.\ 111--229]{Bir1950}.  
\bibitem
{Mather84}
\Author{J. Mather}
\ArticleTitle{Nonexistence of invariant circles} 
\JournalTitle{Ergodic Theory Dynam. Systems} 
{\bf 4} (1984), 301-309, reprinted in 
\cite[pp. 395--403]{MacKayM87}.  
\bibitem
{MatherF94}
\Author{J. J. Mather and G. Forni}
\ArticleTitle{Action minimizing orbits in Hamiltonian systems}
\BookTitle{Transition to Chaos in Classical and Quantum Mechanics
          (Montecatini Terme, 1991)}
	J. Bellissard et al, Eds., 
          Springer, Berlin, (1994), 92--186.
\bibitem
{Khin97}
\Author{A. Ya. Khinchin}
\BookTitle{Continued Fractions}
Dover, Mineola, NY, 1997.
\bibitem
{MacKayP87}
\Author{R. S. MacKay and I. C. Percival}
\ArticleTitle{Converse KAM: theory and practice}
\JournalTitle{Comm. Math. Phys.} {\bf 98} (1985), 469--512.
\bibitem
{Jungreis91}
\Author{I. Jungreis}
\ArticleTitle{A method for proving that monotone twist maps have no invariant
          circles}
\JournalTitle{Ergodic Theory Dynam. Systems} {\bf 11} (1991), 79--84.
\bibitem
{DeV58}
\Author{R. DeVogelaere} 
\ArticleTitle{On the structure of symmetric 
periodic solutions of conservative systems, 
with applications}
In 
\JournalTitle{Contributions to the Theory of Nonlinear 
Oscillations}, S. Lefschetz (Ed.), 
Princeton University Press, 1958, pp. 53--84.  
\bibitem
{Greene79}
\Author{J.~M.~Greene}
\ArticleTitle{A method for determining a stochastic transition}
\JournalTitle{J. Math. Phys.} {\bf 20} 
(1979), 1183--1201, 
reprinted in \cite[pp.\ 419--437]{MacKayM87}.  
\bibitem
{ApteLP05}
\Author{A. Apte, R. de la Llave and N. P. Petrov}
\ArticleTitle{Regularity of critical invariant circles 
of the standard nontwist map}  
\JournalTitle{Nonlinearity} {\bf 18} (2005), 1173--1187.
\bibitem
{Ketoja92}
\Author{J. A. Ketoja}
\ArticleTitle{Renormalisation in a circle map with two inflection points}
\JournalTitle{Phys. D} {\bf 55} 
(1992), 45--68. 
\bibitem
{KetojaM94}
\Author{J. A. Ketoja and R. S. MacKay}
\ArticleTitle{Rotationally-ordered periodic orbits for multiharmonic
     area-preserving twist maps}
\JournalTitle{Phys. D} {\bf 73} (1994), 388--398.
\bibitem
{FalcoliniL92-domains}
\Author{C. Falcolini and R. de la Llave}
\ArticleTitle{Numerical calculation of domains of analyticity
   perturbation theories in the presence of small divisors}
\JournalTitle{J. Stat. Phys.} {\bf 67} (1992), 645--666.
\bibitem
{LlaveO06}
\Author{R. de la Llave and A. Olvera}
\ArticleTitle{The obstruction criterion for non-existence of invariant 
     circles and renormalization}
\JournalTitle{Nonlinearity} {\bf 19} (2006) 1907--1937.  
\bibitem
{LlaveS96}
\Author{R. de la Llave and R. P. Schafer}
\ArticleTitle{Rigidity properties of one dimensional expanding maps
  and applications to renormalization}
Manuscript, 1996.
\bibitem
{Birkhoff25}
\Author{G. D. Birkhoff}
\ArticleTitle{An extension of Poincar\'e's last geometric theorem} 
\JournalTitle{Acta Math.} {\bf 47} (1925), 297--311, 
reprinted in \cite[pp.\ 252--266]{Bir1950}.  
\bibitem
{KH}
\Author{A.~Katok and B.~Hasselblatt}
\BookTitle{Introduction to the Modern Theory of Dynamical Systems} 
Cambridge University Press, Cambridge, 1995.
\bibitem
{OS87}
\Author{A. Olvera and C. Sim\'o}
\ArticleTitle{An obstruction method for the 
  destruction of invariant curves}
\JournalTitle{Phys. D} {\bf 26} (1987),  181--192.
\bibitem
{FalcoliniL92-Greene}
\Author{C. Falcolini and R. de la Llave}
\ArticleTitle{A rigorous partial justification of Greene's criterion} 
\JournalTitle{J. Stat. Phys.} {\bf 67} (1992), 609--643.
\bibitem
{Carletti03}
\Author{T. Carletti}
\ArticleTitle{The $1/2$-complex Bruno function and the Yoccoz function: 
	a numerical study of the Marmi-Moussa-Yoccoz conjecture}
\JournalTitle{Experiment. Math.} {\bf 12} 
(2003), 491--506.
\bibitem
{Stein70}
\Author{E.~Stein}
\BookTitle{Singular Integrals 
  and Differentiability Properties of Functions}
Princeton University Press, Princeton, 1970.
\bibitem
{Stirnemann97}
\Author{A. Stirnemann}
\ArticleTitle{Towards an existence proof of MacKay's fixed point}
\JournalTitle{Commun. Math. Phys.}
{\bf 188} (1997), 723--735.
\bibitem
{HentschelP83}
\Author{H. G. E. Hentschel and I. Procaccia}
\ArticleTitle{The infinite number of generalized dimensions 
of fractals and strange attractors}
\JournalTitle{Phys. D} {\bf 8} (1983), 435--444.
\bibitem
{Renyi60}
\Author{A. R\'enyi}
\ArticleTitle{On measures of entropy and information}
in 
\BookTitle{Proceedings of the Fourth Berkeley Symposium 
on Mathematical Statistics and Probability, University of California, 
June 20--July 30, 1960, Vol~I}
University of California Press, Berkeley, 1961, 
pp.~547--561.  
\bibitem
{Jaffard97}
\Author{S. Jaffard}
\ArticleTitle{Multifractal formalism for functions. 
I. Results valid for all functions. 
II. Self-similar functions}
\JournalTitle{SIAM J. Math. Anal.} 
{\bf 28} 
(1997), 944--970, 971--998. 
\bibitem
{OsbaldestinS87}
\Author{A. H. Osbaldestin and M. Y. Sarkis}
\ArticleTitle{Singularity spectrum of a critical {KAM} torus}
\JournalTitle{J. Phys. A} {\bf 20} (1987), L953--L958.
\bibitem
{BuricMT97}
\Author{N. Buri\'c, M. Mudrini\'c and K. Todorovi\'c}
\ArticleTitle{Equivalent classes of critical circles}
\JournalTitle{J. Phys. A} {\bf 30} (1997), L161--L165.
\bibitem
{BuricMT98}
\Author{N. Buri\'c, M. Mudrini\'c and K. Todorovi\'c}
\ArticleTitle{Universal scaling of critical quasiperiodic orbits in a class
        of twist maps}
\JournalTitle{J. Phys. A} {\bf 39} (1998), 7848--7854.
\bibitem
{ShiH91}
\Author{J. Shi and B. Hu}
\ArticleTitle{Crossover phenomena in the multifractal behavior 
	of invariant circles}
\JournalTitle{Phys. Lett. A} {\bf 156} (1991), 267--271.
\bibitem
{HuS94}
\Author{B. Hu and J. Shi}
\ArticleTitle{Nonanalytic twist maps and Frenkel-Kontorova model}
\JournalTitle{Phys. D} {\bf 71} (1994), 23--38.
\bibitem
{HuntKSY96}
\Author{B. R. Hunt, K. M. Khanin, Ya. G. Sinai, and J. A. Yorke}
\ArticleTitle{Fractal properties of critical invariant curves}
\JournalTitle{J. Stat. Phys.} {\bf 85} (1996), 261--276.  
\bibitem
{VulSK84}
\Author{E. B. Vul, Ya. G. Sinai, and K. M. Khanin}
\ArticleTitle{Feigenbaum universality and thermodynamic formalism}
\JournalTitle{Russian Math. Surveys} {\bf 39} 
(1984), 1--40, 
reprinted in \cite[pp.\ 491--530]{Cvitanovic89}.  
\bibitem
{YamaguchiT99}
\Author{Y. Yamaguchi and K. Tanikawa}
\ArticleTitle{A remark on the smoothness of critical KAM curves 
in the standard mapping}
\JournalTitle{Prog. Theor. Phys.}
{\bf 101} 
(1999), 1--24.
\bibitem
{ArneodoBM95}
\Author{A. Arneodo, E. Bacry and J. F. Muzy} 
\ArticleTitle{The thermodynamics of fractals revisited 
with wavelets}
\JournalTitle{Phys. A} {\bf 213} (1995), 232--275.
\bibitem
{MuzyBA94}
\Author{J. F. Muzy, E. Bacry and A. Arneodo} 
\ArticleTitle{The multifractal formalism revisited with wavelets}
\JournalTitle{Internat. J. Bifur. Chaos Appl. Sci. Engrg.} 
{\bf 4} 
(1994), 245--302.  
\bibitem
{Grace}
\Author{The Grace team}
Grace homepage.  
{\tt http://plasma-gate.weizmann.ac.il/Grace/}.  
\bibitem
{Cvitanovic89}
\Author{P. Cvitanovi\'c}
\BookTitle{Universality in Chaos}
IOP Publishing, Bristol, 1989.  
\bibitem
{Bir1950}
\Author{G. D. Birkhoff}
\BookTitle{Collected Works, Vol II} 
Dover, New York, 1968.  
\bibitem
{MacKayM87}
\Author{R. S. MacKay and J. D. Meiss}
\BookTitle{Hamiltonian Dynamical Systems} 
Adam Hilger, Bristol, 1993.







\pagebreak


\section*{Tables}


\vfill


\begin{table}[h]
\begin{center}
\begin{tabular}{ccccc}
\hline
\hline
$F$ with: & $\regul(R)$ & $\regul(g)$ & $\regul(h)$ & $\regul(h^{-1})$ \\
\hline
\hline
$V_1$  & $1.83(9)$ & $1.83(9)$ & $0.722(1)$ & $0.92(1)$ \\
\hline
$V_2$  & $1.79(6)$ & $1.75(9)$ & $0.721(1)$ & $0.92(1)$ \\
\hline
$V_3$  & $1.83(4)$ & $1.84(3)$ & $0.724(2)$ & $0.93(2)$ \\
\hline
$V_4$  & $1.86(8)$ & $1.86(8)$ & $0.722(1)$ & $0.92(1)$ \\
\hline
$V_5$  & $1.85(5)$ & $1.85(5)$ & $0.724(2)$ & $0.93(1)$ \\
\hline
$V_6$  & $1.85(15)$ & $1.88(12)$ & $0.726(3)$ & $0.93(2)$ \\
\hline
\hline
\end{tabular}
\caption{\label{table:regul-ghR}Regularities 
of the functions $R$, $g$, $h$, and~$h^{-1}$ 
for the golden critical invariant circles of 
different maps~$F$.}  
\end{center}
\end{table}


\vfill





\pagebreak


\section*{Figure captions}


\noindent
{\bf Caption to Figure \ref{fig:R}:} \\[0.5mm]
Critical invariant circles, $r=R(\theta)$, 
of the maps corresponding to the maps $V_1$, $V_2$, $\ldots$, 
$V_6$ given by \eqref{eq:armo_1}--\eqref{eq:trian} 
($V_1$ = thin solid line, 
$V_2$ = thick solid line, 
$V_3$ = dotted line, 
$V_4$ = thin dashed line, 
$V_5$ = thick dashed line, 
$V_6$ = dot-dashed line).

\vspace{3mm}

\noindent
{\bf Caption to Figure \ref{fig:g}:} \\[0.5mm]
''Periodized'' advance maps $g-\mathrm{Id}$ 
(notation same as in Fig.~\ref{fig:R}).

\vspace{3mm}

\noindent
{\bf Caption to Figure \ref{fig:h-shifted}:} \\[0.5mm]
``Periodized'' conjugacies $h-\mathrm{Id}$ 
(notation same as in Fig.~\ref{fig:R}).

\vspace{3mm}

\noindent
{\bf Caption to Figure \ref{fig:all-hm1}:} \\[0.5mm]
``Periodized'' inverse conjugacies $h^{-1}-\mathrm{Id}$ 
(notation same as in Fig.~\ref{fig:R}).

\vspace{3mm}

\noindent
{\bf Caption to Figure \ref{fig:zoom}:} \\[0.5mm]
Zooming in the graph 
of the function $h-\mathrm{Id}$ 
corresponding to the map $V_2$ \eqref{eq:armo_101}.


\vspace{3mm}

\noindent
{\bf Caption to Figure \ref{fig:bigH}:} \\[0.5mm]
``Periodized'' big conjugacies~$H-\mathrm{Id}$.

\vspace{3mm}

\noindent
{\bf Caption to Figure \ref{fig:spectrum-h}:} \\[0.5mm]
Plot of $\log_{10}\left|\left(\widehat{h-\mathrm{Id}}\right)_k\right|$ 
versus $\log_{10} k$, 
where $h$ corresponds to the map $F$ 
coming form the function $V_3$~\eqref{eq:two}.  

\vspace{3mm}

\noindent
{\bf Caption to Figure \ref{fig:spectrum-two-g-hm1}:} \\[0.5mm]
Plot of 
$\log_{10}\left|\left(\widehat{g-\mathrm{Id}}\right)_k\right|$ 
and 
$\log_{10}\left|\left(\widehat{h^{-1}-\mathrm{Id}}\right)_k\right|$ 
versus $\log_{10} k$, 
for the same map $F$ as in Fig.~\ref{fig:spectrum-h}.  
The impulses correspond to $(g-\mathrm{Id})$, 
and the dots above them to $(h^{-1}-\mathrm{Id})$.  

\vspace{3mm}

\noindent
{\bf Caption to Figure \ref{fig:clp-slopes}:} \\[0.5mm]
Plots of 
$\log_{10}\left\| \left( \frac{\partial}{\partial t}\right)^{\eta} 
{\rm e}^{-t\sqrt{-\Delta}} K \right\|_{L^{\infty}(\bbT)}$
versus $\log_{10}t$ for the functions $K=(h-\mathrm{Id})$ 
for the twist maps coming from 
$V_1$, $\ldots$, $V_6$, 
for $\eta=2$ (shallowest lines), $\eta=3$, 
and $\eta=4$ (steepest lines).  

\vspace{3mm}

\noindent
{\bf Caption to Figure \ref{fig:clp-diffs}:} \\[0.5mm]
Slope of the lines on Fig.~\ref{fig:clp-slopes} 
as a function of~$\log_{10}t$ (see the text). 
The notation is the same as in Fig.~\ref{fig:R}.  






\pagebreak


\section*{Figures}


\vfill


\begin{figure}[h]
\centerline{
\epsfig{file=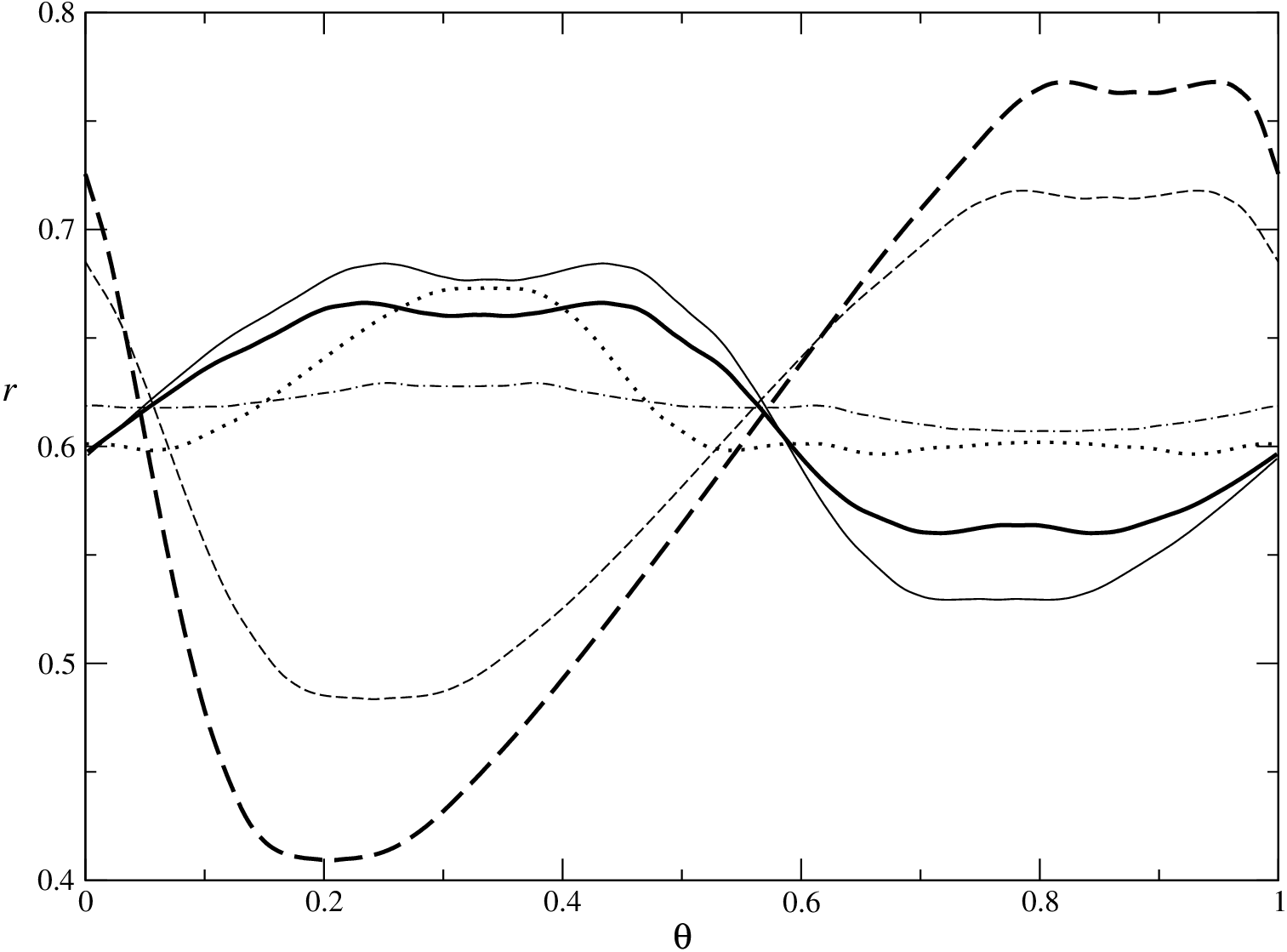, width=1.0\textwidth,angle=0}
}
\caption{\label{fig:R}
}
\end{figure}

\vfill


\pagebreak


\begin{figure}[bpt]
\centerline{
\epsfig{file=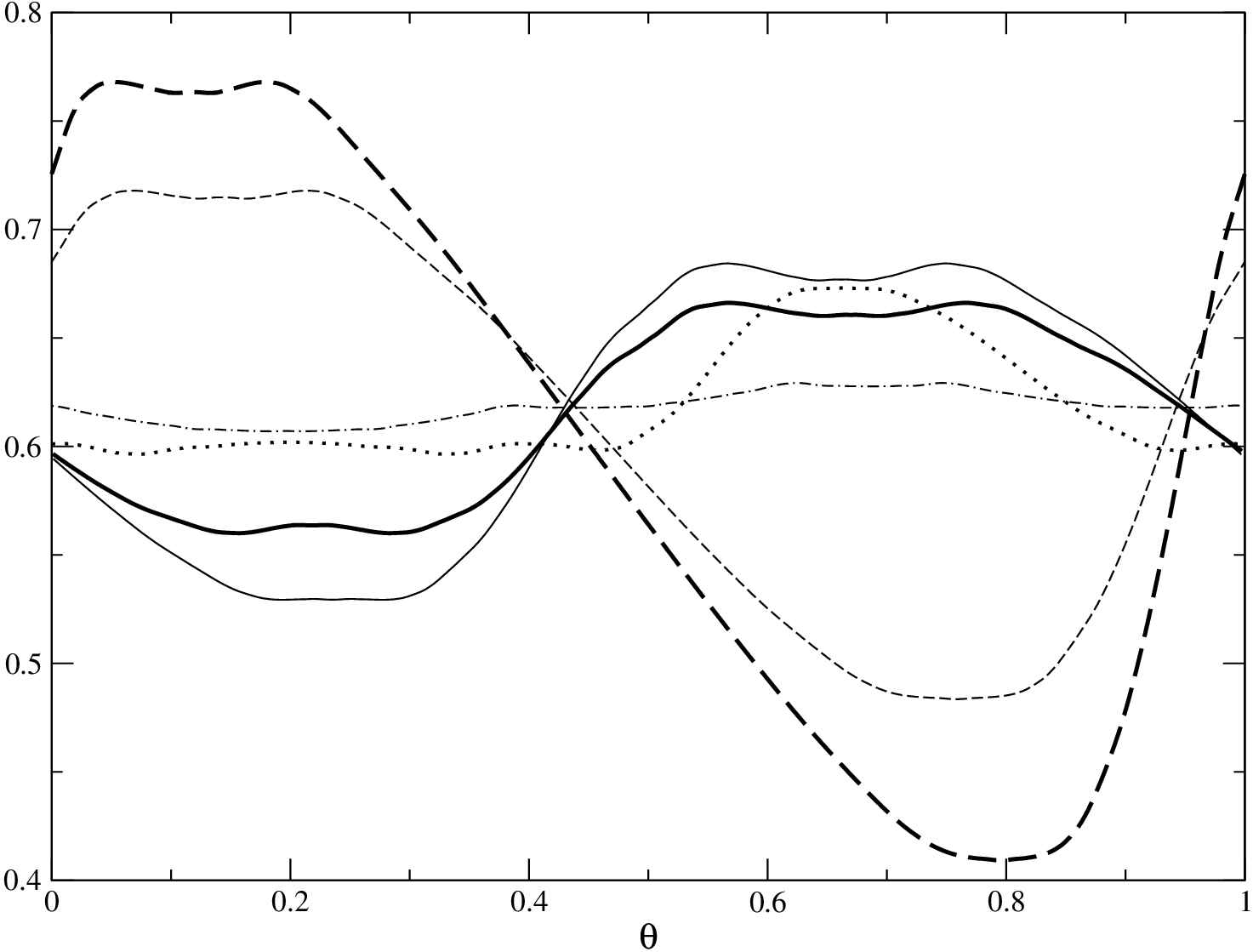, width=1.0\textwidth,angle=0}
}
\caption{\label{fig:g}
}
\end{figure}


\begin{figure}[bpt]
\centerline{
\epsfig{file=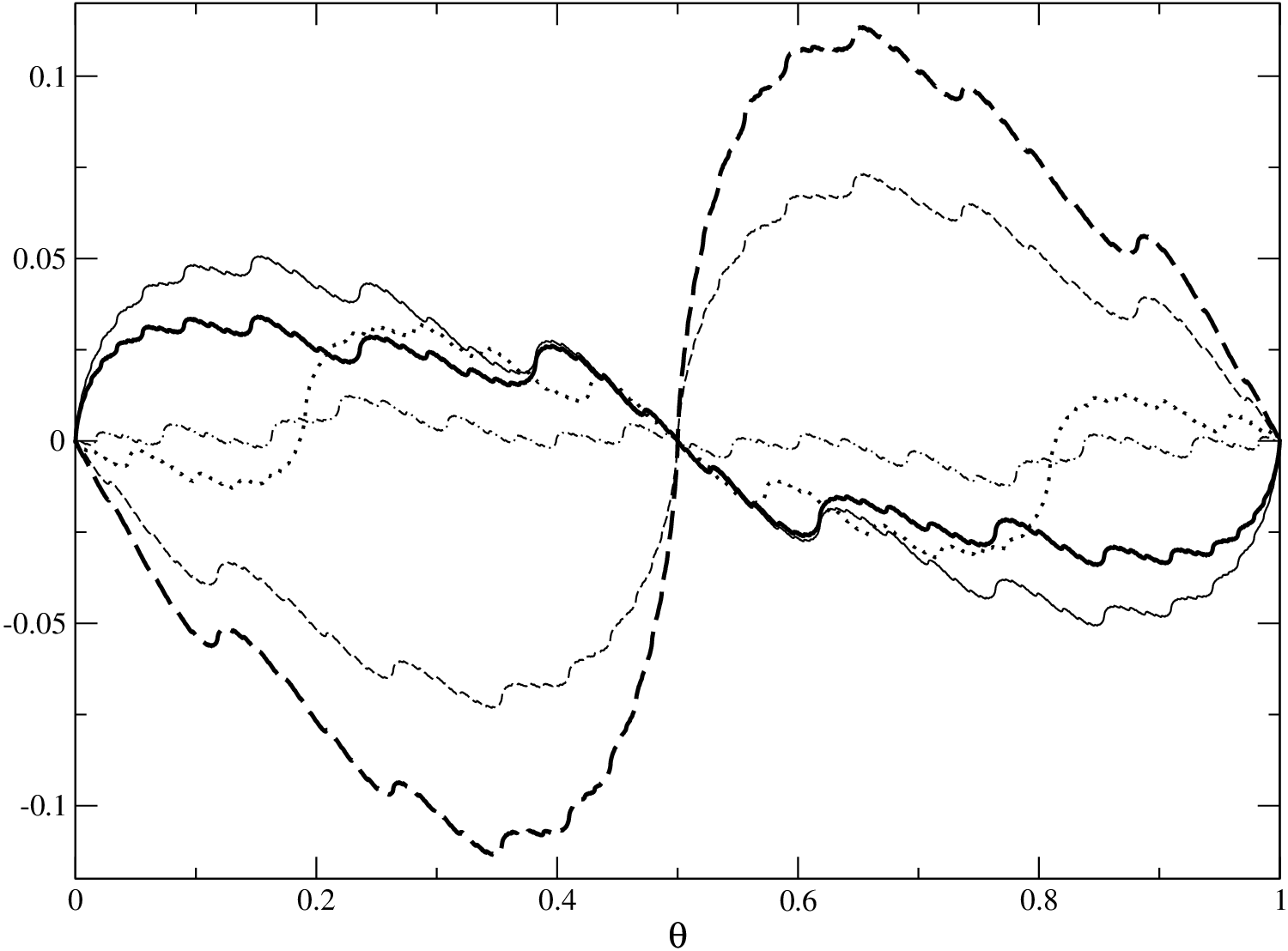, width=1.0\textwidth,angle=0}
}
\caption{\label{fig:h-shifted}
}
\end{figure}


\begin{figure}[bpt]
\centerline{
\epsfig{file=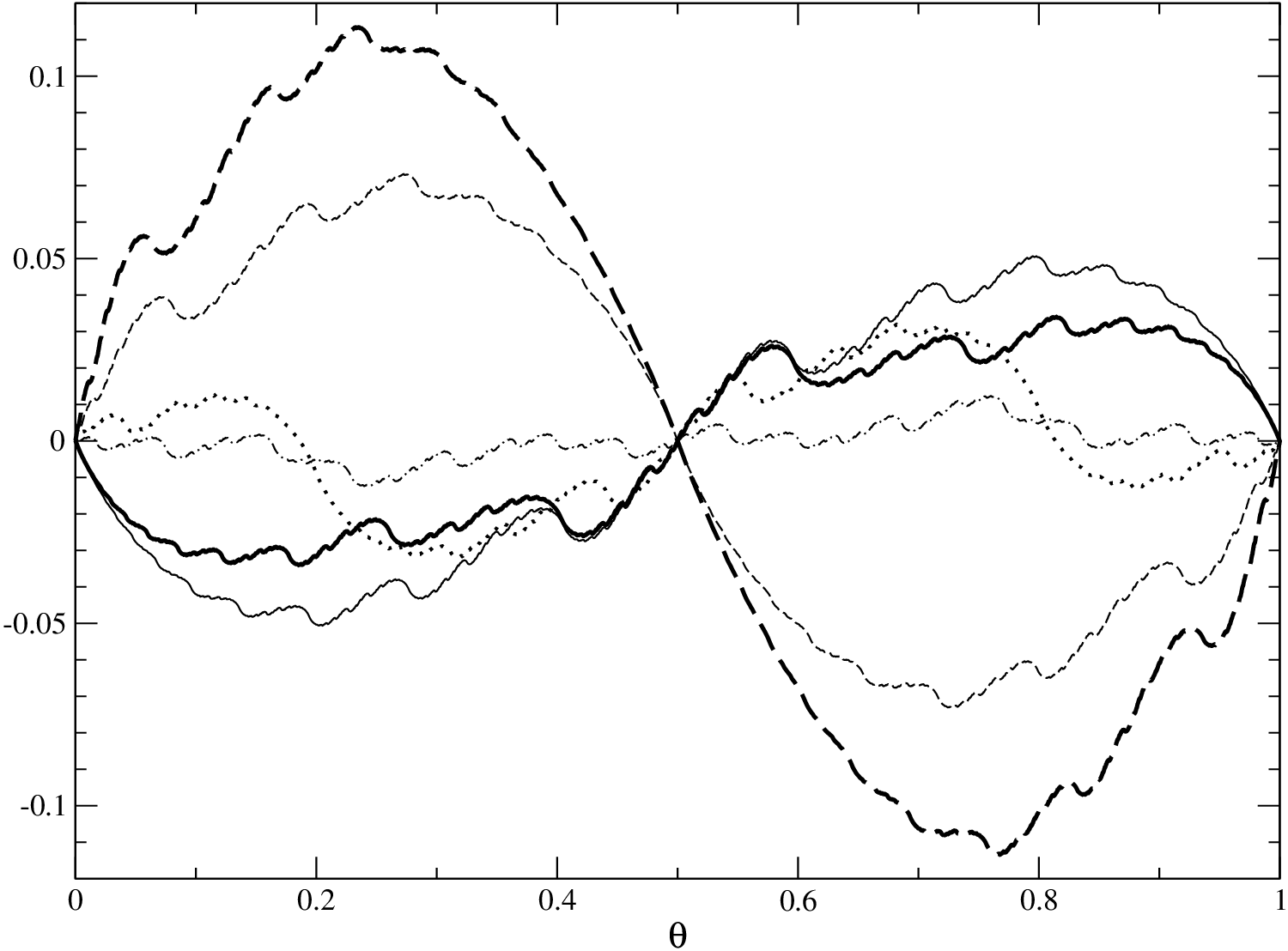, width=1.0\textwidth,angle=0}
}
\caption{\label{fig:all-hm1}
}
\end{figure}



\begin{figure}[bpt]
\centerline{
\epsfig{file=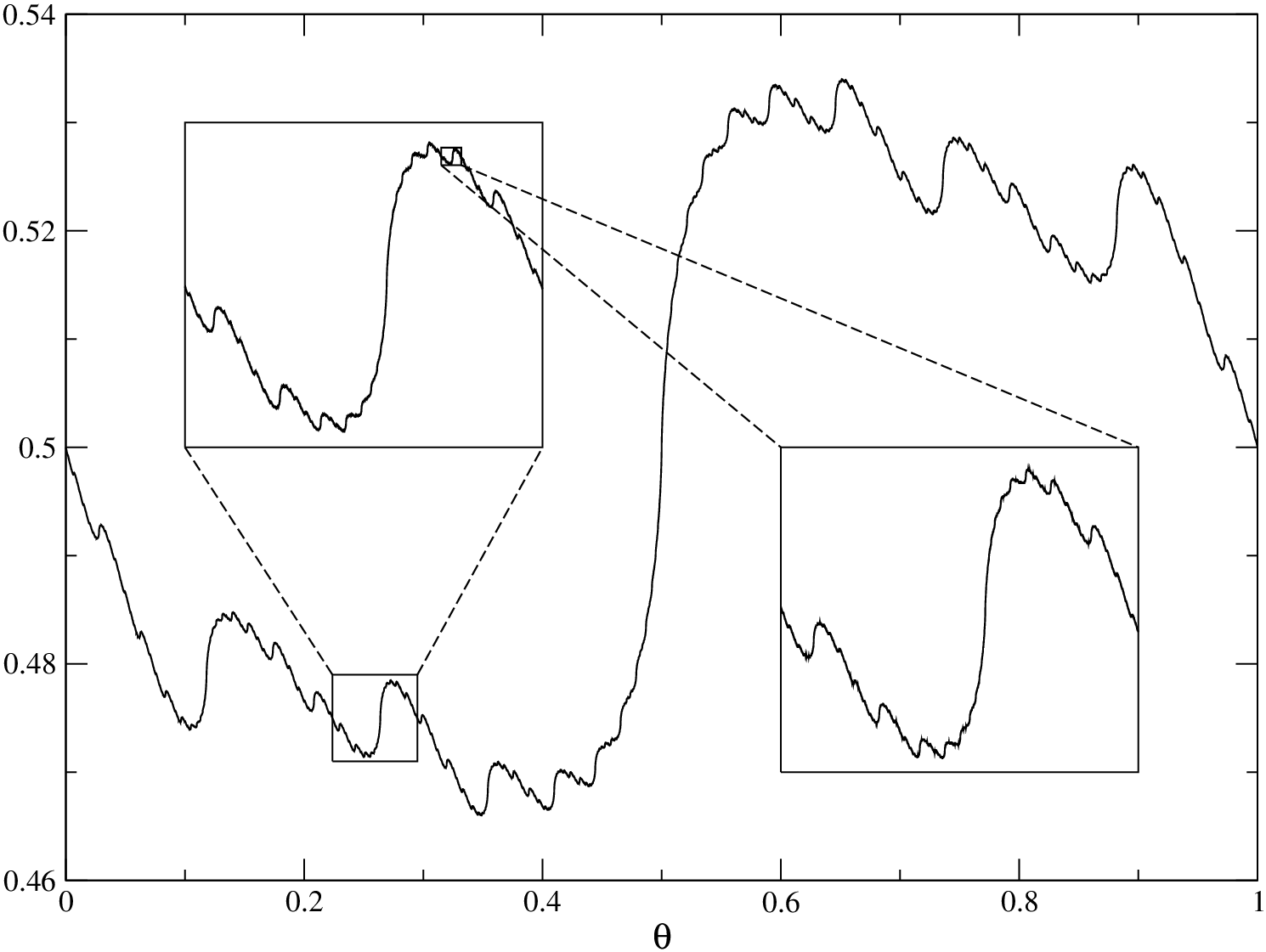, width=1.0\textwidth,angle=0}
}
\caption{\label{fig:zoom}
}
\end{figure}


\begin{figure}[bpt]
\centerline{
\epsfig{file=bigH.eps, width=1.0\textwidth,angle=0}
}
\caption{\label{fig:bigH}
}  
\end{figure}


\begin{figure}[bpt]
\centerline{
\epsfig{file=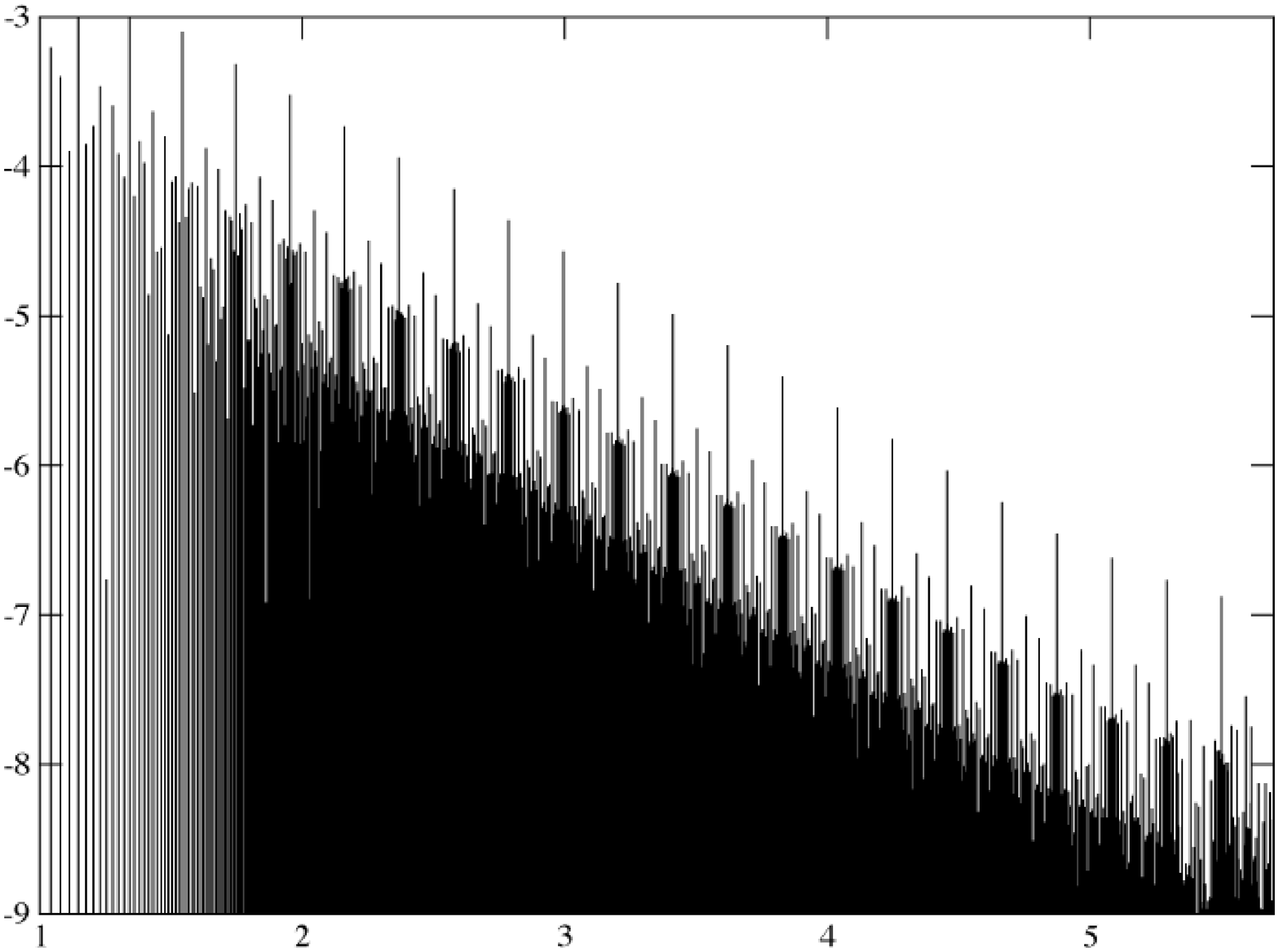, width=1.0\textwidth,angle=0}
}
\caption{\label{fig:spectrum-h}
}
\end{figure}


\begin{figure}[bpt]
\centerline{
\epsfig{file=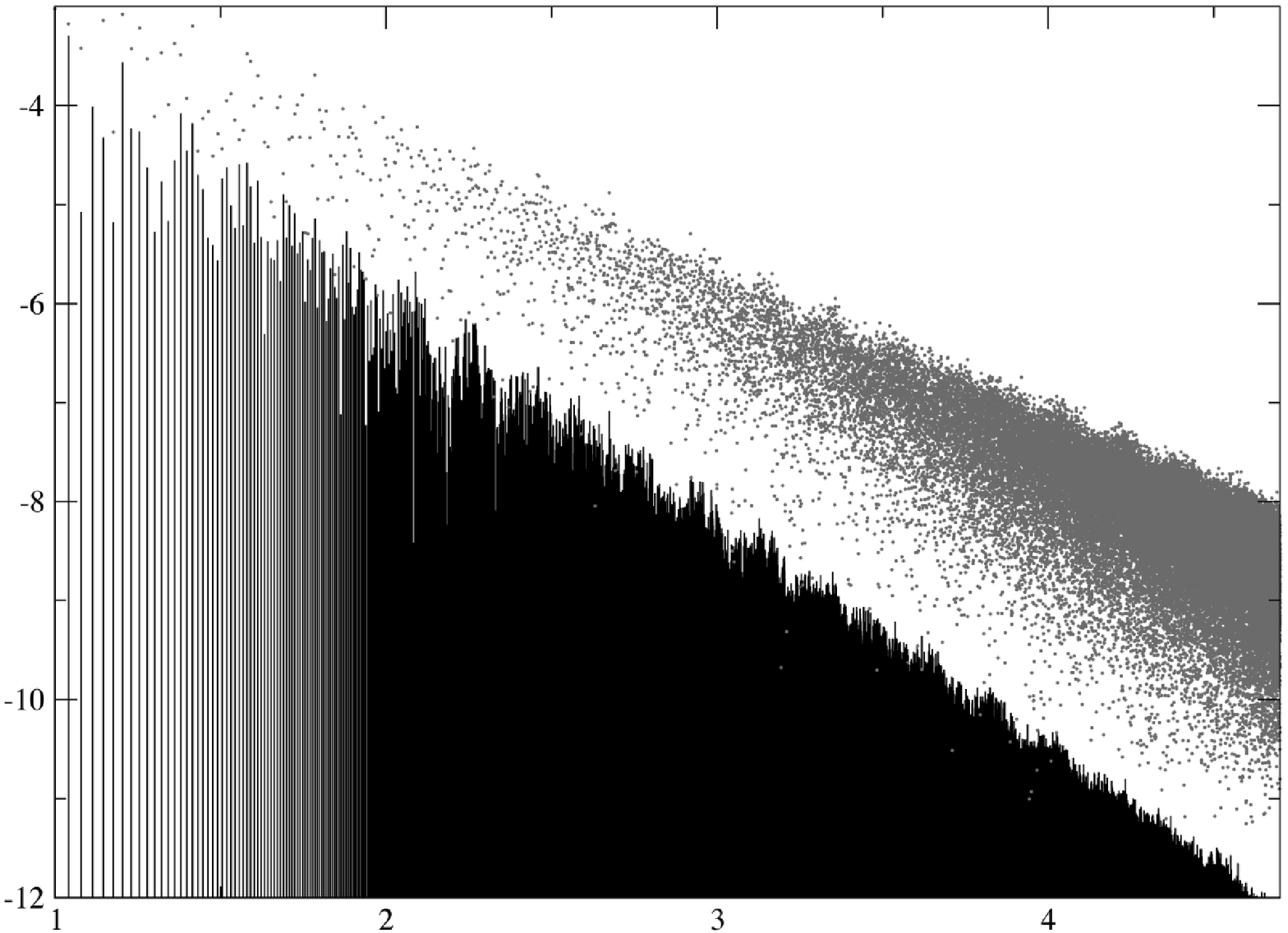, width=1.0\textwidth,angle=0}
}
\caption{\label{fig:spectrum-two-g-hm1}
}
\end{figure}


\begin{figure}[bpt]
\centerline{
\epsfig{file=clp-slopes.eps, width=1.0\textwidth,angle=0}
}
\caption{\label{fig:clp-slopes}
}
\end{figure}


\begin{figure}[bpt]
\centerline{
\epsfig{file=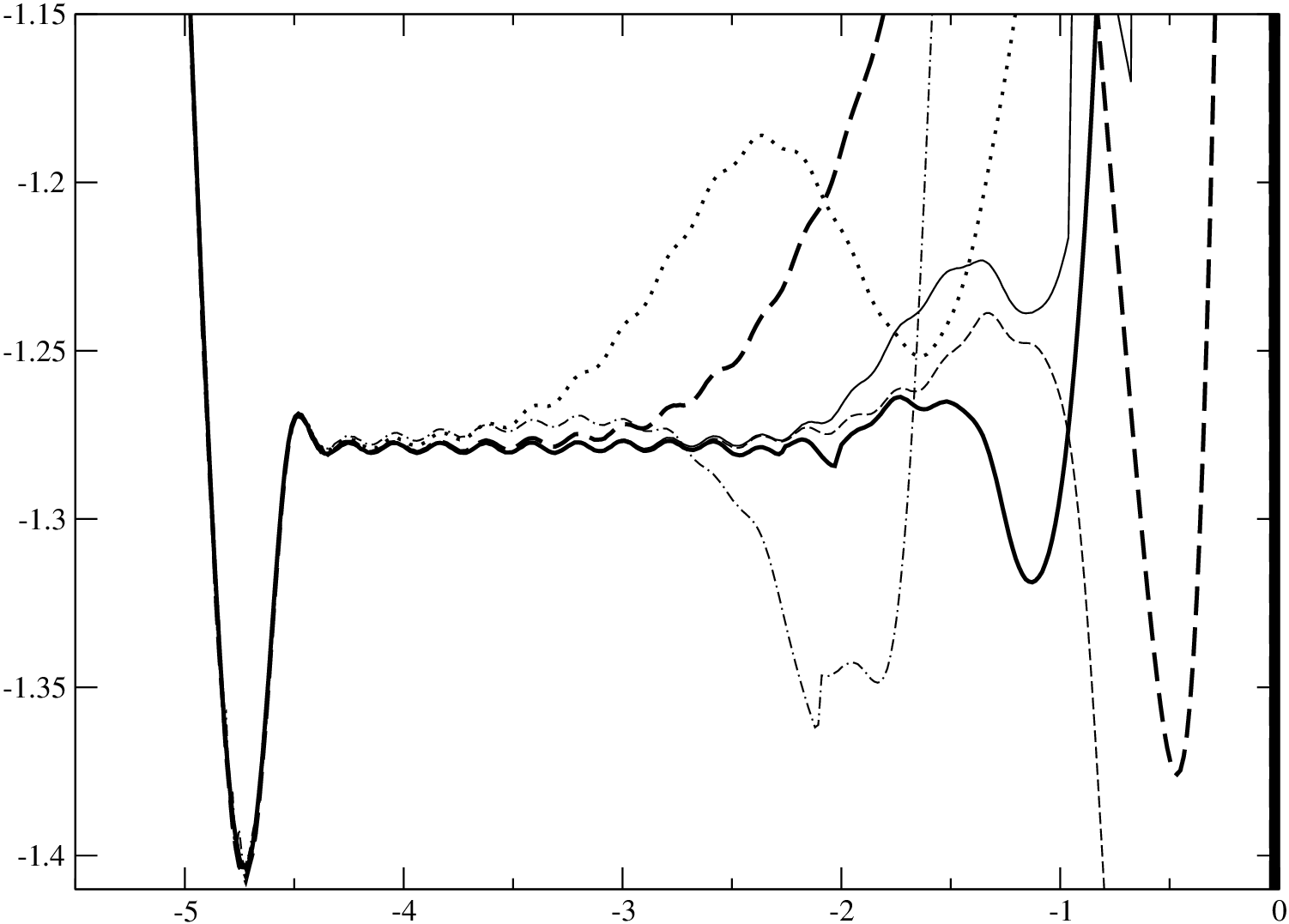, width=1.0\textwidth,angle=0}
}
\caption{\label{fig:clp-diffs}
}
\end{figure}






%
%
%
%
%








%
%
%
%

\end{thebibliography}




\end{document}